%% file: fb.tex
\documentclass[11pt, a4paper]{article}

\usepackage{jcappub}

\usepackage{graphicx}
\usepackage{datetime}
\usepackage{textpos}
\usepackage{booktabs}
\newcommand{\kpc}{\ \text{kpc}}
\newcommand{\GeV}{\ \text{GeV}}
\newcommand{\s}{\ \text{s}}
\newcommand{\cm}{\ \text{cm}}
\newcommand{\fex}{\textit{e.g.}~}
\newcommand{\cf}{\textit{cf.}~}

\title{A Tentative Gamma-Ray Line\\ from Dark Matter Annihilation\\ at the
Fermi Large Area Telescope}

\author{Christoph Weniger}
\affiliation{Max-Planck-Institut f\"ur Physik, F\"ohringer Ring 6, 80805
M\"unchen, Germany}
\emailAdd{weniger@mppmu.mpg.de}

\abstract{The observation of a gamma-ray line in the cosmic-ray fluxes would
be a smoking-gun signature for dark matter annihilation or decay in the
Universe.  We present an improved search for such signatures in the data of
the Fermi Large Area Telescope (LAT), concentrating on energies between $20$
and $300\GeV$. Besides updating to 43 months of data, we use a new data-driven
technique to select optimized target regions depending on the profile of the
Galactic dark matter halo. In regions close to the Galactic center, we find a
$4.6\sigma$ indication for a gamma-ray line at $E_\gamma\approx130\GeV$. When
taking into account the look-elsewhere effect the significance of the observed
excess is~$3.2\sigma$. If interpreted in terms of dark matter particles
annihilating into a photon pair, the observations imply a dark matter mass of
$m_\chi=129.8\pm 2.4\,^{+7}_{-13}\GeV$ and a partial annihilation
cross-section of $\langle\sigma v\rangle_{\chi\chi\to\gamma\gamma}
=\left(1.27\pm0.32\,^{+0.18}_{-0.28}\right)\times 10^{-27}\cm^3\s^{-1}$ when
using the Einasto dark matter profile. The evidence for the signal is based on
about $50$ photons; it will take a few years of additional data to clarify its
existence.}

\begin{document}
\begin{textblock}{2}(10.1,0)
  \noindent
  MPP-2012-73
\end{textblock}

\maketitle

\section{Introduction}
The existence of dark matter (DM) in the Universe is a by now well established
fact, supported by numerous of observations that range from the time of
matter/radiation equality until today~\cite{Komatsu:2010fb}. A plethora of
possible particle physics scenarios for DM exists, among which the currently
leading hypothesis are weakly interacting massive particles (WIMPs) that are
thermally produced in the early Universe~\cite{Jungman:1995df, Bertone:2004pz,
Bergstrom:2009ib}.  WIMPs could be produced at particle colliders like the
CERN LHC, they could be observed via WIMP-nucleon scattering in low-background
experiments, and, due to their self-annihilation in the Universe, they could
contribute to the cosmic-ray fluxes (see Ref.~\cite{Cirelli:2012tf} for a
recent review). 

Among the different particle species produced in WIMP annihilation,
\emph{gamma rays} are of particular importance, as they propagate essentially
unperturbed and preserve spatial information about their sources. This
information is exploited for efficient signal/background discrimination in
searches for gamma-ray signatures from \fex dwarf galaxies~\cite{Essig:2009jx,
Scott:2009jn, Abramowski:2010aa, Aleksic:2011jx, Abdo:2010ex,
Ackermann:2011wa, GeringerSameth:2011iw, Cholis:2012am, Mazziotta:2012ux} or
the Galactic center (GC)~\cite{Aharonian:2006wh, Hooper:2011ti, Vitale:2011zz,
Abramowski:2011hc}.

Besides spatial signatures, signatures in the \emph{energy spectrum} of the
gamma-ray flux play a major role in DM searches. It was noted long
ago~\cite{Bergstrom:1988fp} that the two-body annihilation of DM into photons
produces monochromatic gamma rays that could stand out of the otherwise
continuous astrophysical fluxes as a clear smoking gun
signature~\cite{Rudaz:1989ij, Bergstrom:1997fh, Ullio:1997ke, Bern:1997ng, Bergstrom:2004nr,
Bertone:2009cb, Bertone:2010fn}.  In most cases, the line signal is expected
to be extremely faint, as its production is one-loop suppressed.  However, it
can be enhanced in many scenarios, like non-minimal variations of the
MSSM~\cite{Profumo:2008yg, Ferrer:2006hy}, singlet DM~\cite{Profumo:2010kp},
hidden U(1) DM~\cite{Dudas:2009uq, Mambrini:2009ad}, effective DM
scenarios~\cite{Goodman:2010qn}, scenarios with strong annihilation into the
higgs boson and a photon~\cite{Jackson:2009kg} or inert higgs doublet
DM~\cite{Gustafsson:2007pc}; these gamma-ray lines would be indeed in reach of
current technology~\cite{Bringmann:2011ye}. Additional prominent spectral
features in the gamma-ray flux are generated by the emission of internal
bremsstrahlung (IB) photons during DM annihilation~\cite{Beacom:2004pe,
Bergstrom:2004cy, Birkedal:2005ep, Bergstrom:2005ss, Bringmann:2007nk}.

Since more than three years, the Fermi Large Area Telescope (LAT), which is
the main instrument on the Fermi Gamma-ray Space
Telescope~\cite{Atwood:2009ez}, measures the gamma-ray sky with an
unprecedented precision, searching---amongst others---for DM
signals~\cite{Baltz:2008wd}. Previous searches for spectral signatures from DM
annihilation~\cite{Abdo:2010nc, Vertongen:2011mu, Fermi:Symp2011Talk,
Edmonds:2011PhD, Feng:2011ab} were exclusively concentrating on gamma-ray lines (an earlier
analysis using EGRET data can be found in Ref.~\cite{Pullen:2006sy}).  A first
dedicated search for IB signatures was recently presented in
Ref.~\cite{Bringmann:2012vr}; in this paper, a week indication for a spectral
signature close to the GC was reported.

The purpose of the present work is to update and refine the existing searches
for gamma-ray lines from DM annihilation in Fermi LAT data. In particular, we
are interested in a further investigation of the signature reported in
Ref.~\cite{Bringmann:2012vr}.  Besides taking into account all data collected
so far (43 months), and making use of the most up-to-date publicly available
event selections, we employ a new data-driven algorithm to find target regions
that maximize the signal-to-noise ratio (SNR) for different profiles of the
Galactic DM halo.  Since we are looking for very faint signatures, this is
extremely important.  An inefficient choice of the target region can easily
hide a DM signal. We concentrate on line energies between $20$ and $300\GeV$
and use the data below $20\GeV$ for target region selection. 

The rest of this paper is organized as follows: In Section~\ref{sec:analysis},
we explain in detail the DM signal we are looking for, the treatment of the
Fermi LAT data, the target region selection and the spectral analysis; in
Section~\ref{sec:results} we present our main results; a discussion of
instrumental systematics, further tests and comparison with previous work can
be found in Section~\ref{sec:caveats}; we conclude in
Section~\ref{sec:conclusions}. Event tables are provided in
Appendix~\ref{apx:tables}.

\section{Data Analysis}
\label{sec:analysis}
\subsection{DM annihilation in the Galactic halo}
The gamma-ray flux from DM particles~$\chi$ annihilating into a photon pair
inside the Galactic DM halo is (for a Majorana WIMP) given by
\begin{equation}
  \frac{dJ_\gamma}{dEd\Omega}(\xi) = \frac{\langle \sigma
  v\rangle_{\chi\chi\to\gamma\gamma}}{8\pi \,m_\chi^2} \, 2\delta(E-E_\gamma)
  \int_\text{l.o.s.}\!\!\!\!\! ds\;\rho_\text{dm}^2(r)\;,
  \label{eqn:fluxADM}
\end{equation}
where $\xi$ is the angle to the GC. Here, $m_\chi$ is the DM mass, $\langle
\sigma v\rangle_{\chi\chi\to\gamma\gamma}$ the partial annihilation
cross-section for $\chi\chi\to\gamma\gamma$, $E_\gamma=m_\chi$ the gamma-ray
line energy and $\rho_\text{dm}(r)$ the DM distribution as function of the
Galactocentric distance $r$.  The coordinate $s\geq0$ runs along the
line-of-sight, and $r(s,\xi) = \sqrt{(r_0-s\cos\xi)^2 + (s\sin\xi)^2}$, where
$r_0=8.5\kpc$ denotes the distance between Sun and GC.  \medskip

We consider five reference DM profiles, all normalized to
$\rho_\text{dm}(r_0)=0.4\GeV\cm^{-3}$ at position of the Sun~\cite{Catena:2009mf,
Salucci:2010qr}: (1) The Einasto profile, favored by the latest $N$-body
simulations~\cite{Navarro:2003ew, Springel:2008cc, Pieri:2009je},
\begin{align}
  \rho_{\text{dm}}(r) \propto \exp \left(-\frac{2}{\alpha_E}
  \frac{r^{\alpha_E} }{r_s^{\alpha_E}}\right)\;,
\end{align}
with $\alpha_E=0.17$ and $r_s=20\kpc$. (2) The commonly used cored
isothermal profile
\begin{align}
  \rho_\text{dm}(r) \propto \frac{1}{
  1+\left(r/r_s\right)^2}\;, \label{}
\end{align}
with $r_s=3.5\kpc$ (see \fex Ref.~\cite{Salucci:2007tm} for observational
arguments in favor of cored DM profiles). (3-5) The generalized
Navarro-Frenk-White (NFW) profile
\begin{align}
  \rho_\text{dm}(r) \propto \frac{1}{(r/r_s)^\alpha
  \left(1+r/r_s\right)^{3-\alpha}}\;,
  \label{eqn:NFW}
\end{align}
with $r_s=20\kpc$. Here, $\alpha$ parametrizes the profile's inner slope, and
$\alpha=1$ reproduces the usual NFW profile~\cite{Navarro:1996gj,
Abdo:2010nc}. The presence of the super-massive black hole~\cite{Gondolo:1999ef}
or adiabatic contraction~\cite{Blumenthal:1985qy, Gnedin:2003rj,
Gustafsson:2006gr, Gnedin:2011uj} can steepen the inner profile, which is here
effectively taken into account by allowing that $\alpha>1$. To this end, we
consider the three reference values $\alpha=(1.0,\,1.15,\,1.3)$, which are
compatible with microlensing and dynamical observations~\cite{Iocco:2011jz}.
Note that changing the slope of the inner profile in principle also affects
the profile normalization. Since we are here mainly interested in the shape of
the DM signal, however, we will keep $\rho_\text{dm}(r_0)=0.4\GeV\cm^{-3}$
fixed for simplicity.

\subsection{Data selection}
In the present analysis, we take into account 43 months of data (from 4 Aug
2008 to 8 Mar 2012) with energies between 1 and 300 GeV.\footnote{High level
data of the LAT is available at \url{http://fermi.gsfc.nasa.gov}.} We apply
the zenith-angle cut $\theta<100^\circ$ in order to avoid contamination with
the earth albedo, as well as the recommended quality-filter cut DATA\_QUAL==1.
For comparison and cross-checks, we make use of both the SOURCE and ULTRACLEAN
events selections (both Pass 7 Version 6). The former features an effective
area that is relatively larger by about $12\%$ at 100 GeV, the latter a lower
contamination with residual charged cosmic rays (CRs). In the main analysis,
front- and back-converted events are considered simultaneously, in order to
minimize statistical errors. The selection of events as well as the
calculation of exposure maps is performed using the 6 Oct 2011 version of
ScienceTools~v9r23p1.\footnote{See
\url{http://fermi.gsfc.nasa.gov/ssc/data/analysis} for the standard chains.}
For everything else we use our own software.

\subsection{Target regions \& observed fluxes}
\begin{figure}
  \centering
  \includegraphics[width=0.44\linewidth]{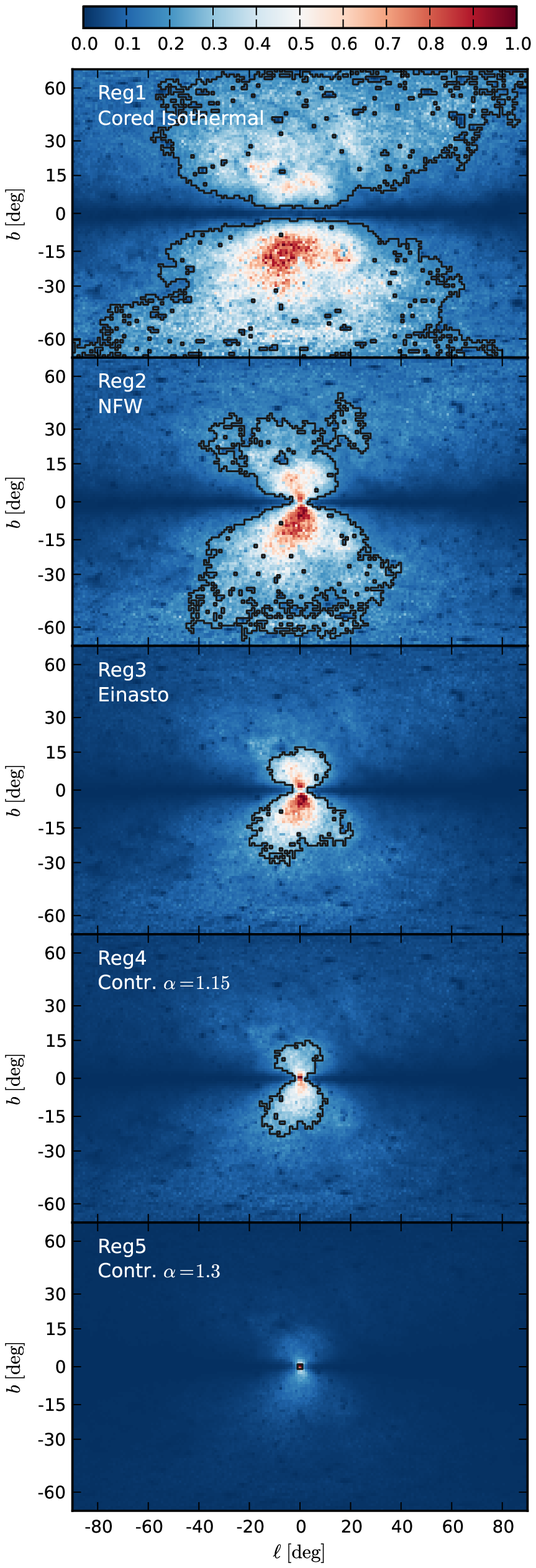}
  \includegraphics[width=0.49\linewidth]{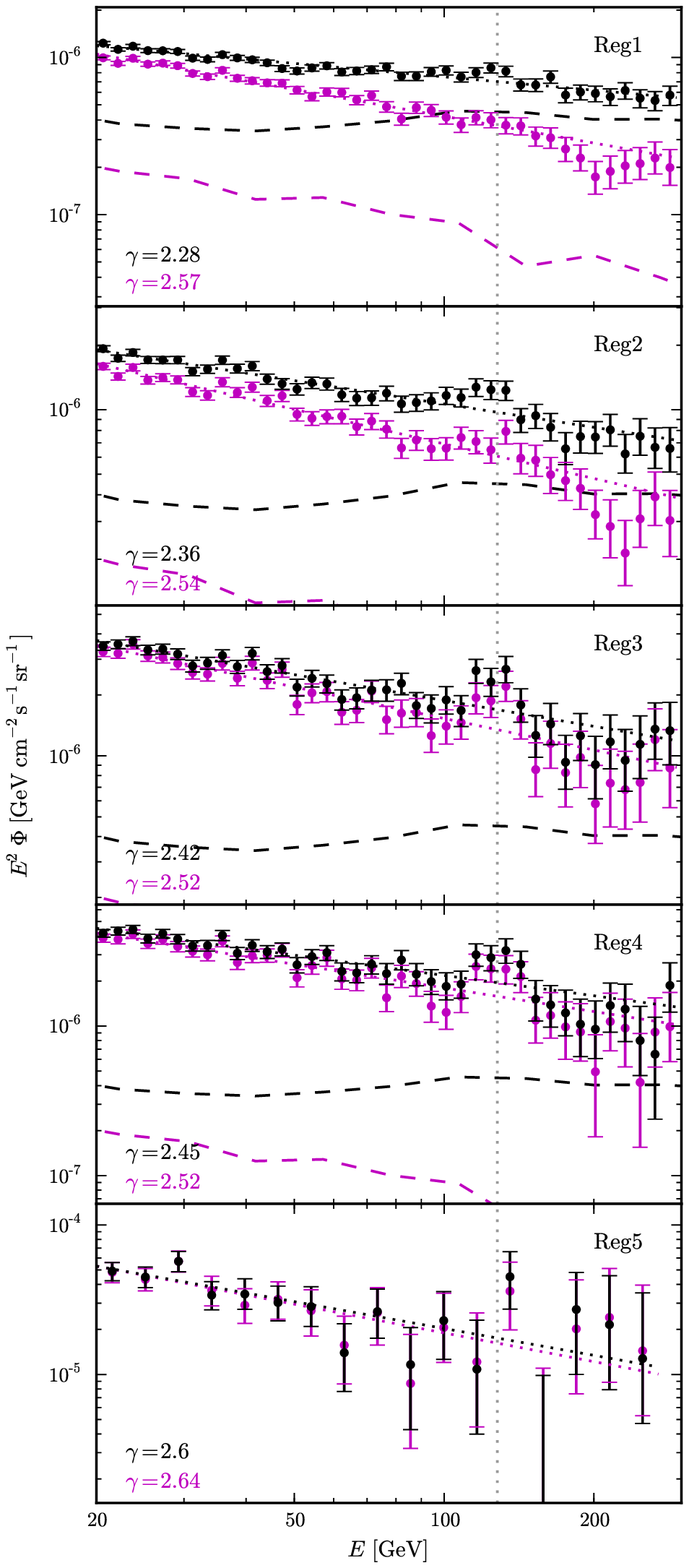}
  \vspace{-0.4cm}
  \caption{\emph{Left panel:} The \emph{black lines} show the target regions
  that are used in the present analysis in case of the SOURCE event class (the
  ULTRACLEAN regions are very similar). From \emph{top} to \emph{bottom}, they
  are respectively optimized for the cored isothermal, the NFW (with
  $\alpha=1$), the Einasto and the contracted (with $\alpha=1.15$, $1.3$) DM
  profiles. The \emph{colors} indicate the signal-to-\emph{background} ratio
  with arbitrary but common normalization; in Reg2 to Reg5 they are
  respectively downscaled by factors (1.6, 3.0, 4.3, 18.8) for better
  visibility.  \newline
  \emph{Right panel:} From top to bottom, the panels show the $20$--$300\GeV$
  gamma-ray (+ residual CR) spectra as observed in Reg1 to Reg5 with
  statistical error bars. The SOURCE and ULTRACLEAN events are shown in
  \emph{black} and \emph{magenta}, respectively. \emph{Dotted lines} show
  power-laws with the indicated slopes; \emph{dashed lines}
  show the EGBG + residual CRs. The \emph{vertical gray} line indicates
  $E=129.0\GeV$.}
  \label{fig:reg}
\end{figure}

The optimal target region for gamma-ray line searches maximizes the SNR and
depends on both, the morphology of the DM signal and the morphology of the
background flux. For a particular Galactic DM profile, the former is given by
Eq.~\eqref{eqn:fluxADM}; the latter has to be in principle determined from a
modeling of gamma-ray diffuse and point sources at $20$--$300\GeV$ energies.
The basic strategy that we follow here is to approximate the background
morphology above 20 GeV by the spatial distribution of gamma rays measured
between $1$ and $20\GeV$.  

From events in the energy range $1$--$20\GeV$, we produce a two-dimensional
count map that covers an area of $|b|<84^\circ$ Galactic latitude and
$|\ell|<90^\circ$ Galactic longitude, using a cylindrical equal-area
projection with one square-degree pixel size.\footnote{Because of this
projection we exclude data with $|b|>84^\circ$. This does not affect the
results. Note that we use the convention $\ell=-180^\circ\dots180^\circ$
throughout.}  For each pixel $i$, we derive the number of expected signal events
$\mu_i$ from Eq.~\eqref{eqn:fluxADM}, whereas the number of actually measured
events is denoted by $c_i$.  We need $\mu_i$ only up to an overall
normalization, since we leave $\langle\sigma
v\rangle_{\chi\chi\to\gamma\gamma}$ and $m_\chi$ unspecified at this point.
As long as the signal is only a weak perturbation, the SNR in each pixel can
be estimated by $\mathcal{R}_i=\mu_i/\sqrt{c_i}$. Note that throughout the
entire analysis, the angular resolution of the LAT---$\Delta\theta\approx
0.2^\circ$ for $20$ to $300\GeV$ energies~\cite{Fermi:performancePASS7}---is
neglected.

We define the \emph{optimal} target region as the set of pixels
$\mathcal{T}_o$ for which the integrated SNR
\begin{align}
  \mathcal{R}_{\mathcal{T}_o}=\frac{\sum_{i\in\mathcal{T}_o} \mu_i }{
  \sqrt{\sum_{i\in\mathcal{T}_o} c_i}}\;,
\end{align}
is maximized. To find $\mathcal{T}_o$, we use the approximate but efficient
algorithm from Ref.~\cite{Bringmann:2012vr}. The resulting target regions,
optimized for our five reference DM profiles, are shown in the left panels of
Fig.~\ref{fig:reg} by the black lines. The colors indicate the
signal-to-\emph{background} ratio $\mu_i/c_i$, with an arbitrary but common
normalization. In case of a cored isothermal profile (Reg1), the target region
is largest and reaches up to latitudes of $|b|\simeq84^\circ$. The smallest
region (Reg5) corresponds to a compressed profile with inner slope
$\alpha=1.3$, and contains the central $2^\circ\times2^\circ$ degree of the GC
only. In most cases, the regions are more extended south from the GC. This is
a consequence of a slight north/south asymmetry in the observed diffuse
gamma-ray flux (see \fex~Ref.~\cite{FermiLAT:2012aa}). Note that the regions
are only optimal as long as signal contributions are small. In the presence of
a potential signal these regions can be further optimized, which we leave for
future work.

We extract from the LAT data the gamma-ray flux measured in each of the five
target regions. The corresponding energy spectra between $20$ and $300\GeV$
are shown in the right panels of Fig.~\ref{fig:reg} for both, SOURCE (black)
and ULTRACLEAN (magenta) event classes. The residual CR contamination of the
SOURCE event selection is best visible in Reg1 as a sizeable difference
between the SOURCE and ULTRACLEAN fluxes.  This is further illustrated by the
dashed lines, which show the expected flux of residual CRs plus the
extragalactic gamma-ray background (EGBG) for
comparison~\cite{Fermi:BackgroundModels}.  Remarkably, in Reg3 and Reg4 a
pronounced bump at energies around $130\GeV$ (indicated by the vertical dotted
line) can be easily recognized by eye; this spectral feature will turn out to
be the best candidate for a gamma-ray line in the Fermi LAT data between $20$
and $300\GeV$.

\subsection{Spectral analysis} 
\begin{figure}
  \begin{center}
    \includegraphics[width=0.6\linewidth]{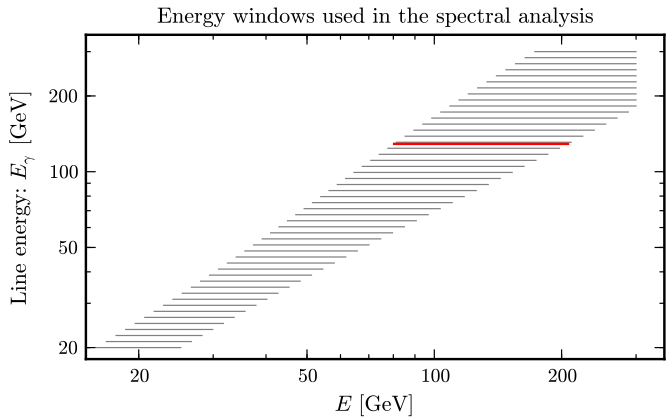}
  \end{center}
  \caption{Energy windows that we use for our spectral line search. In
  \emph{red} we indicate the window that enters the fit at
  $E_\gamma=129.0\GeV$.}
  \label{fig:ewinPos}
\end{figure}

In order to search for gamma-ray lines in Reg1 to Reg5, we perform a shape
analysis of the energy spectra shown in Fig.~\ref{fig:reg} (though with much
smaller energy bins). For a given gamma-ray line energy $E_\gamma$, this
analysis is done in a small energy window that contains $E_\gamma$. The exact
positions of the energy windows adopted during the main analysis are shown in
Fig.~\ref{fig:ewinPos}.  Since the energy windows follow the gamma-ray line
energy, this method is known as ``sliding energy window''
technique~\cite{Pullen:2006sy, Abdo:2010nc, Vertongen:2011mu,
Bringmann:2011ye, Bringmann:2012vr}. 

We can conveniently parametrize the boundaries of the energy window as
\begin{align}
  E_0=E_\gamma/\sqrt{\epsilon} \quad \text{and}\quad
  E_1=\min(E_\gamma\sqrt{\epsilon},\ 300\GeV)\;.
  \label{eqn:ewin}
\end{align}
The sizes of the adopted energy windows vary between $\epsilon \simeq 1.6$ at
low energies, which is a few times wider than the LAT energy resolution, and
$\epsilon\simeq3.0$ at high energies, in order to compensate for the lower
number of events. The choice of the energy window size is somewhat arbitrary,
but depends in principle on the uncertainties in the background curvature, the
effective area and on the available statistics. We will discuss below how a
change of the window size affects the results.

Within the adopted energy windows, we fit the spectra from Fig.~\ref{fig:reg}
with a simple three-parameter model,
\begin{align}
  \frac{dJ}{dE}=S\ \delta(E-E_\gamma) + \beta
  \left(\frac{E}{E_\gamma}\right)^{-\gamma}\;.
  \label{eqn:model}
\end{align}
Background fluxes are here approximated by a single power law with a free
spectral index $\gamma$ and normalization $\beta$, whereas the monochromatic
DM signal has a free normalization $S\geq0$ while its position $E_\gamma$
remains fixed during the fit. Note that, after fixing the experimental
conditions and the profile of the Galactic DM halo, the annihilation
cross-section $\langle\sigma v\rangle_{\chi\chi\to\gamma\gamma}$ is related to
$S$ by a straightforward rescaling.

The best-fit model parameters $(S_\text{bf},\beta_\text{bf},
\gamma_\text{bf})$ are obtained by maximizing the likelihood function
$\mathcal{L}(S, \beta, \gamma) \equiv \Pi_i P(s_i|\nu_i)$, where
$P(s|\nu)\equiv\nu^s e^{-\nu}/s!$ is the Poisson probability distribution
function; here, $s_i$ ($\nu_i$) denotes the number of measured (expected)
events in energy bin $i$.  In general, $\nu_i$ is a function of the model
parameters and calculated by multiplying the above three-parameter model with
the exposure of the target region, and convolving the resulting function with
the effective energy dispersion of the LAT. The details of this calculation
are discussed below in section~\ref{sec:nui}.  In our analysis we use energy
bins that are much smaller than the energy resolution of the LAT, such that
the analysis becomes independent of the actual binning and hence effectively
``unbinned''.\footnote{We use 300 logarithmic bins per energy decade.} 
 
The \emph{significance} of a line signal for a given value of $E_\gamma$ is
derived from the test statistic
\begin{align}
  TS \equiv -2\ln\frac{\mathcal{L}_\text{null}}{\mathcal{L_\text{best}}}\;,
  \label{eqn:TS}
\end{align}
where $\mathcal{L}_\text{best}=\mathcal{L}(S_\text{bf},\beta_\text{bf},
\gamma_\text{bf})$ is the likelihood of a fit \emph{with} DM contribution, and
$\mathcal{L}_\text{null}$ is the likelihood of a fit \emph{without} DM signal
(with $S=0$ fixed during the fit).  In absence of a line-like signature in the
data, the TS is expected to follow a $0.5\chi^2_{k=0}+0.5\chi^2_{k=1}$
distribution.\footnote{This is a consequence of $S\geq0$. The
probability distribution function of $\chi^2_{k=0}$ is $\delta(TS)$.}  Limits
on, as well as statistical errors of, the annihilation cross-section
$\langle\sigma v\rangle_{\chi\chi\to\gamma\gamma}$ are derived using the
profile likelihood method~\cite{Rolke:2004mj}. For instance, a one-sided
$95\%$~CL (two-sided $68.2\%$~CL) limit is obtained by increasing/decreasing
$S$ from $S_\text{bf}$ and profiling over $\beta$ and $\gamma$ until
$-2\ln(\mathcal{L}/\mathcal{L}_\text{best})$ equals 2.71 (1.0). 

In the spectral analysis we scan over a large number of line energies
$E_\gamma$ and over different target regions to find the maximal TS value.
This reduces the statistical significance of any observed excess due to the
well-known look-elsewhere effect. We attribute 6.8 trials over a
$\chi^2_{k=2}$ distribution to the quasi-continuous scan over line energies
from $E_\gamma=20$ to $300\GeV$. This follows from a Monte Carlo analysis as
well as from a subsampling analysis of the LAT data in regions without a DM
signal; details can be found below in Section~\ref{sec:MC}. Furthermore,
we analysed ten different target regions, from which only the five most
interesting ones are shown in this paper.\footnote{The other target regions
correspond to $\alpha=1.05$, 1.1, 1.2 and 1.4 as well as the Fermi Bubble
template from Ref.~\cite{Su:2010qj} (see also Ref.~\cite{Dobler:2009xz}).}
These target regions are partial subsets of each other, but we
conservatively treat them as being statistically independent. However, we do
not attribute trials to the scan over SOURCE and ULTRACLEAN event classes, as
these are obviously strongly correlated.

In summary, we find that the significance of a maximal TS value
$TS_\text{max}$ can in good approximation be derived from $10\times6.8=68$
trials over a $\chi^2_{k=2}$ distribution. In practice, one has to solve
\begin{eqnarray}
  \text{CDF}(\chi^2_{k=2}; TS_\text{max})^{68} = 
  \text{CDF}(\chi^2_{k=1}; \sigma^2)
  \label{eqn:LEE}
\end{eqnarray}
for $\sigma$. Here, $\text{CDF}(\chi^2_k; x)$ is the cumulative distribution
function, which gives the probability to draw a value smaller than $x$ from a
$\chi^2_k$ distribution.

\subsection{Details on the treatment of instrument responses}
\label{sec:nui}
For a given infinitesimal sky region, the full expression for calculating the
expected number of events $\nu_i$ in a certain energy bin $E_i^-$\ldots$E_i^+$
reads
\begin{equation}
  \nu_i = \int_{E_i^-}^{E_i^+} dE 
  \int dE' \int_0^\pi d\theta \sum_{j=f,b} D(E|E', \theta, j) A(E', \theta, j)
  \frac{dT}{d\theta} \frac{dJ}{dE'}\;.
\end{equation}
Here, $\theta$ is the impact angle of photon events with respect to the
instrument axis, $dT/d\theta$ the observational time under this impact angle,
$A(E', \theta, j)$ the effective area, $E'$ the true energy, $j=f,b$ denotes
front- and back-converted events, $E$ the reconstructed energy, and $D(E|E',
\theta, j)$ is the energy dispersion of the LAT. We define the effective
exposure as
\begin{equation}
  X_{eff}(E') \equiv \int_0^\pi d\theta \sum_{j=f,b} A(E', \theta, j) \frac{dT}{d\theta}\;.
\end{equation}
The expected distribution of the conditional parameters $\theta$ and
$j$ for events with energy $E'$ is
\begin{equation}
  P(\theta, j| E')\equiv \frac{A(E', \theta, j) dT/d\theta}{X_{eff}(E')}\;,
\end{equation}
and we define the effective energy dispersion as
\begin{equation}
  D_{eff}(E|E')\equiv \int_0^\pi d\theta \sum_{j=f,b} D(E|E', \theta, j) P(\theta, j,
  E')\;.
\end{equation}
With these definitions, we finally obtain that the number of expected events
$\nu_i$ in energy bin $i$ can be calculated as
\begin{equation}
  \nu_i = \int_{E_i^-}^{E_i^+} dE 
  \int dE' D_{eff}(E|E') X_{eff}(E') \frac{dJ}{dE'}\;.
\end{equation}
We use this expression, together with Eq.~\eqref{eqn:model}, to determine the
number of expected events $\nu_i$ for our likelihood analysis.
\smallskip

The exposure $X_{eff}(E')$ for the different target regions is derived using
the ScienceTools, which automatically integrate over impact angles and front-
and back-converted events according to the flight history of the instrument.
We first generated lifetime cubes via \texttt{gtltcube} for the 43 months of
data (with the options \texttt{dcostheta=0.025}, \texttt{binz=1} and
\texttt{zmax=100}). Then we used \texttt{gtexpcube2} to generate exposure maps
for the instrument response functions (IRFs) \texttt{P7SOURCE\_V6} and
\texttt{P7ULTRACLEAN\_V6}. These maps are integrated over the target regions
to obtain $X_{eff}(E')$; during this step, we weight the exposure maps with
the profile of the expected dark matter signal to obtain correct
signal normalizations.

The parametrization of $D(E|E', \theta, j)$ can be obtained from the
ScienceTools.\footnote{The functional form of the energy dispersion can be
found \fex on
\url{http://fermi.gsfc.nasa.gov/ssc/data/analysis/documentation/Cicerone/Cicerone_LAT_IRFs};
the files in \texttt{refdata/fermi/caldb/CALDB/data/glast/lat/bcf/edisp/} in
ScienceTools~v9r23p1 contain the required parameters.} To determine $P(\theta,
j| E')$, we directly use the distribution of events that enter our analysis.
Integrating over the $-90^\circ<\ell<90^\circ$ region, we find that at
energies of 100 to 300 GeV 55\% of the events are front-, the rest
back-converted. The distribution of impact angles $\theta$ is derived from the
$-90^\circ<\ell<90^\circ$ data above 1 GeV; it peaks at $\theta\approx
40^\circ$ and becomes negligible at $\theta\gtrsim70^\circ$. We approximate
$P(\theta, j| E')$ by adopting these distributions for all energies $E'$ and
for all observational angles $\Omega$.  This is a reasonable approximation for
the following reasons: The ratio of front- and back-converted events is not
strongly energy dependent, above 1 GeV it varies at most by 7\%; since the
energy resolution for front- and back-converted events differs above 1 GeV at
most by 20\%~\cite{Fermi:performancePASS7}, this weak energy dependence is
negligible for our discussion.  By inspecting the lifetime cube for different
sky regions we find that, for the long time interval we consider, all regions
of the sky are observed under a broad distribution of $\theta$ values with
similar shape. However, since the maximal variation of the energy resolution
with the impact angle is about $28\%$~\cite{Fermi:performancePASS7}, one might
consider the above approximation on the $\theta$ distribution as problematic.
We estimate the impact of this approximation by adopting instead the $\theta$
distribution of events within a radius of $10^\circ$ around the galactic
center. We find that the FWHM and $68\%$ containment width of the energy
dispersion changes by less than $1\%$, which is negligible for our purposes.

Note that the approximations for $P(\theta, j| E')$ affect the energy
dispersion \emph{only}, \textit{i.e.}~the exact shape of the line signal, and
not the calculation of the effective area or the power-law fits to the
background, for which energy dispersion can be neglected anyway. For a region
with $10^\circ$ radius around the galactic center, we compared results
obtained with our routines with the results that can be obtained by using
\texttt{gtrspgen} (which is part of ScienceTools and takes fully into account
effects of energy dispersion, but does not allow the use of arbitrary target
regions) and \texttt{XSPEC}~\cite{Arnaud:XSPEC}, and found excellent
agreement.

\section{Main Results}
\label{sec:results}
\begin{figure}
  \centering
  \includegraphics[width=.49\linewidth]{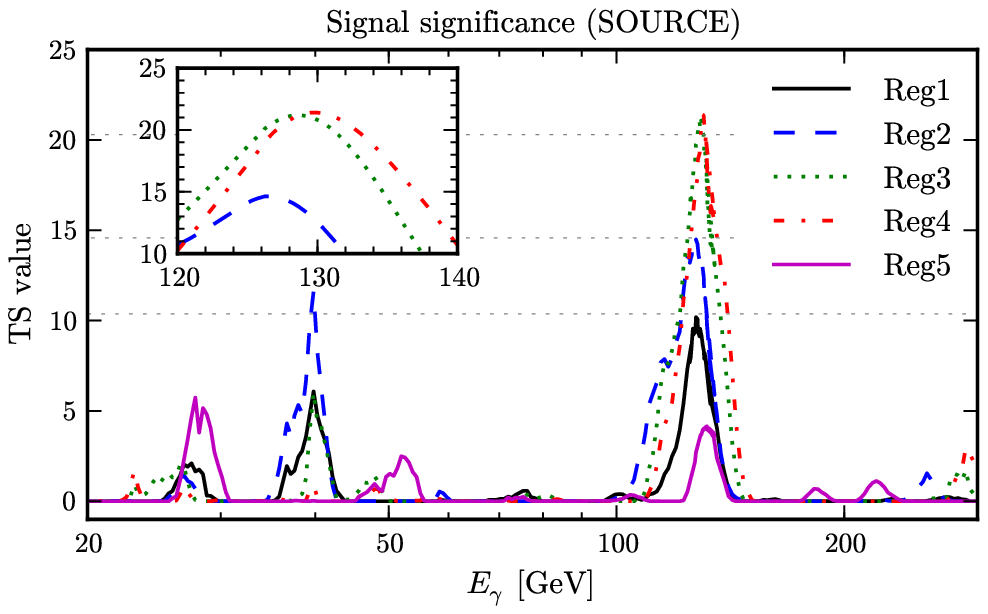}
  \includegraphics[width=.49\linewidth]{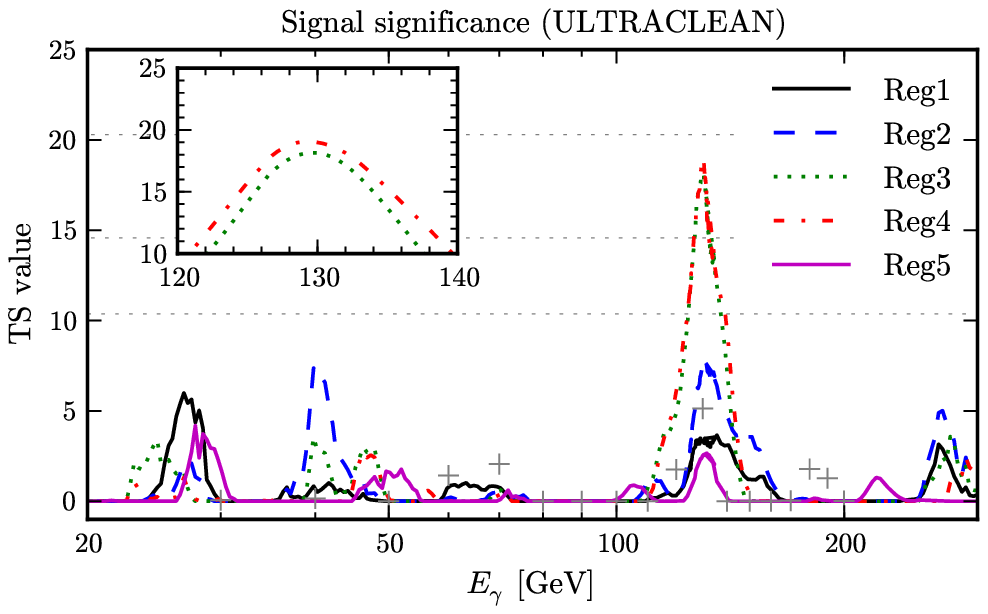}
  \caption{TS value as function of the line energy $E_\gamma$, obtained by
  analysing the energy spectra from the different target regions in
  Fig.~\ref{fig:reg}. \emph{Left} and \emph{right} panels show the results for
  the SOURCE and ULTRACLEAN event classes, respectively.  The inset shows a
  zoom into the relevant region.  The horizontal \emph{gray dotted}
  lines show respectively from bottom to top the $1\sigma$ to 3$\sigma$ levels
  \emph{after} correcting for trials (without trial correction the
  significance is given by $\sqrt{TS}\sigma$). In the right panel, the gray
  crosses show the TS values that we obtain when instead adopting the target
  region and energy windows from Refs.~\cite{Fermi:Symp2011Talk,
  Edmonds:2011PhD} with 43 months of data.}
  \label{fig:sig}
\end{figure}

For each of the spectra shown in Fig.~\ref{fig:reg} we perform a search for
gamma-ray lines in the range $E_\gamma=20$--$300\GeV$ as described above. The
resulting TS values as function of the gamma-ray line energy $E_\gamma$ are
shown in the left and right panels of Fig.~\ref{fig:sig} for the SOURCE and
ULTRACLEAN event classes, respectively.  In regions Reg2, Reg3 and Reg4, we
find TS values that are surprisingly large, and which indicate a high
likelihood for a gamma-ray line at $E_\gamma\approx130\GeV$. The largest TS
value is obtained in case of the SOURCE events in Reg4 and reads $TS=21.4$
(corresponding to $4.6\sigma$ before trial correction). Taking into account
the look-elsewhere effect as discussed above, the trial corrected statistical
significance for the presence of a line signal in the LAT data is $3.2\sigma$.

\begin{figure}
  \centering
  \includegraphics[width=0.46\linewidth]{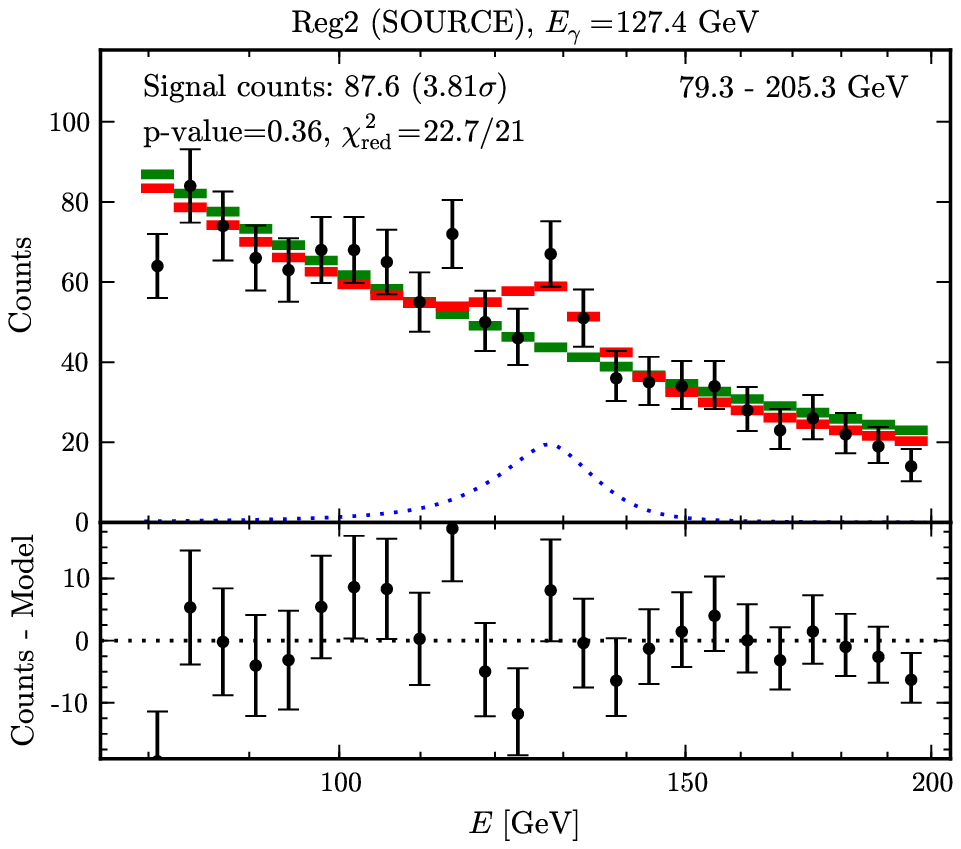}
  \includegraphics[width=0.46\linewidth]{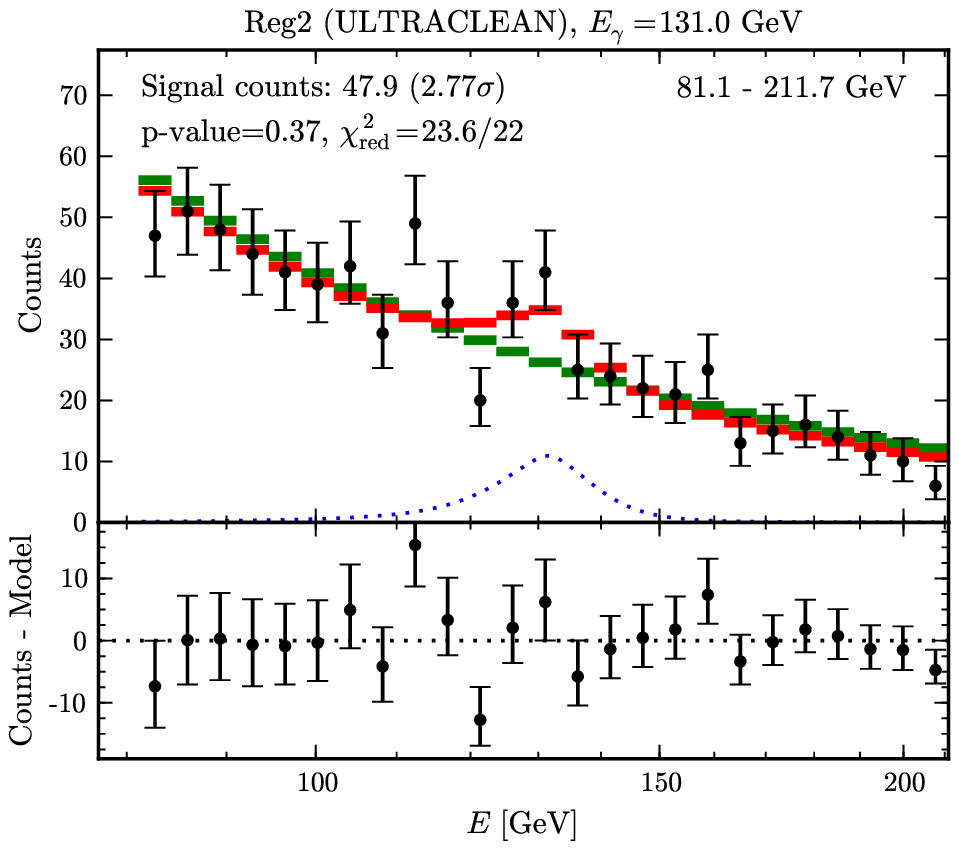}
  \includegraphics[width=0.46\linewidth]{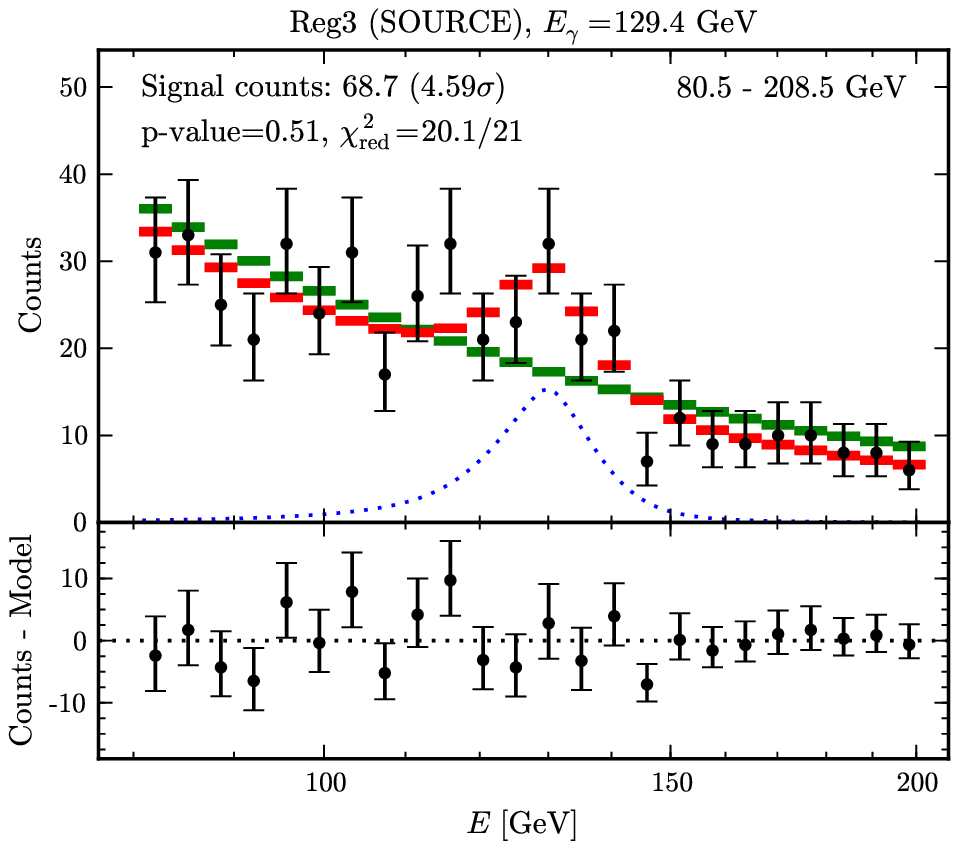}
  \includegraphics[width=0.46\linewidth]{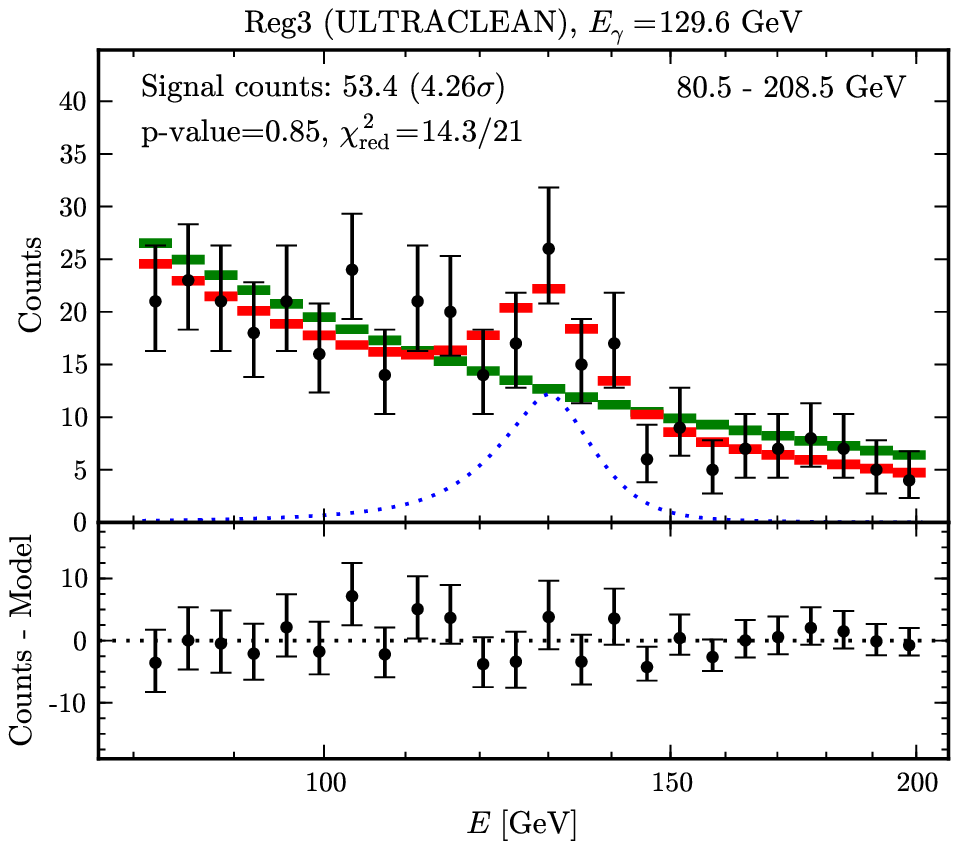}
  \includegraphics[width=0.46\linewidth]{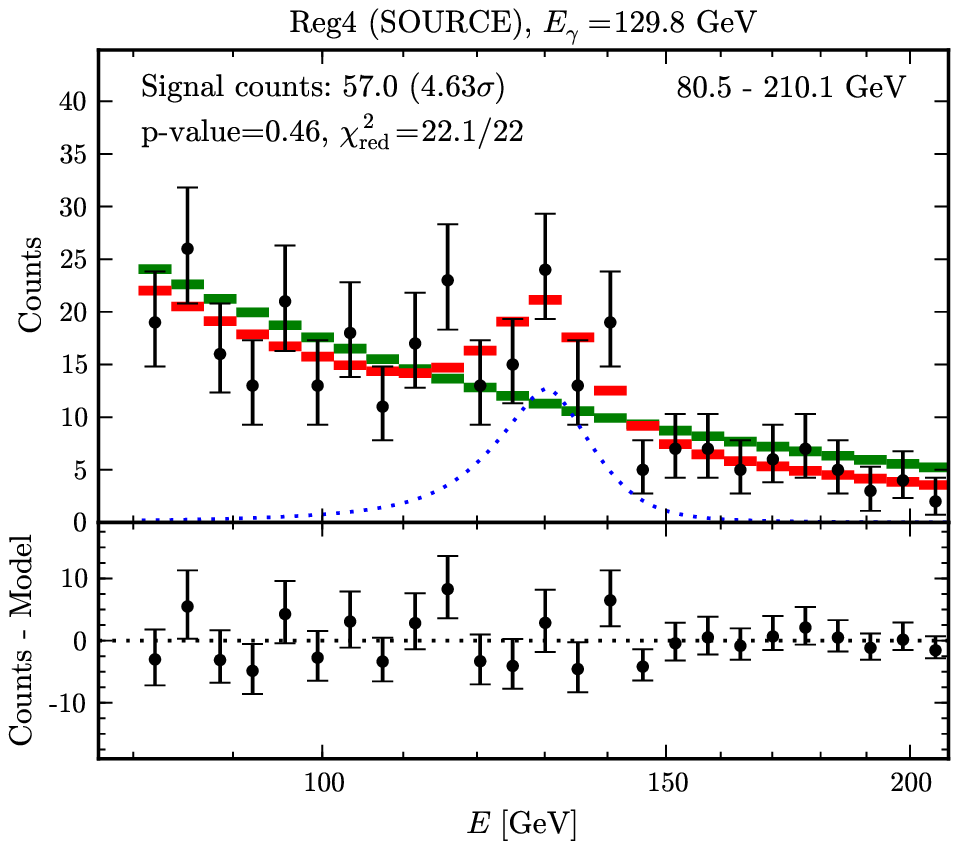}
  \includegraphics[width=0.46\linewidth]{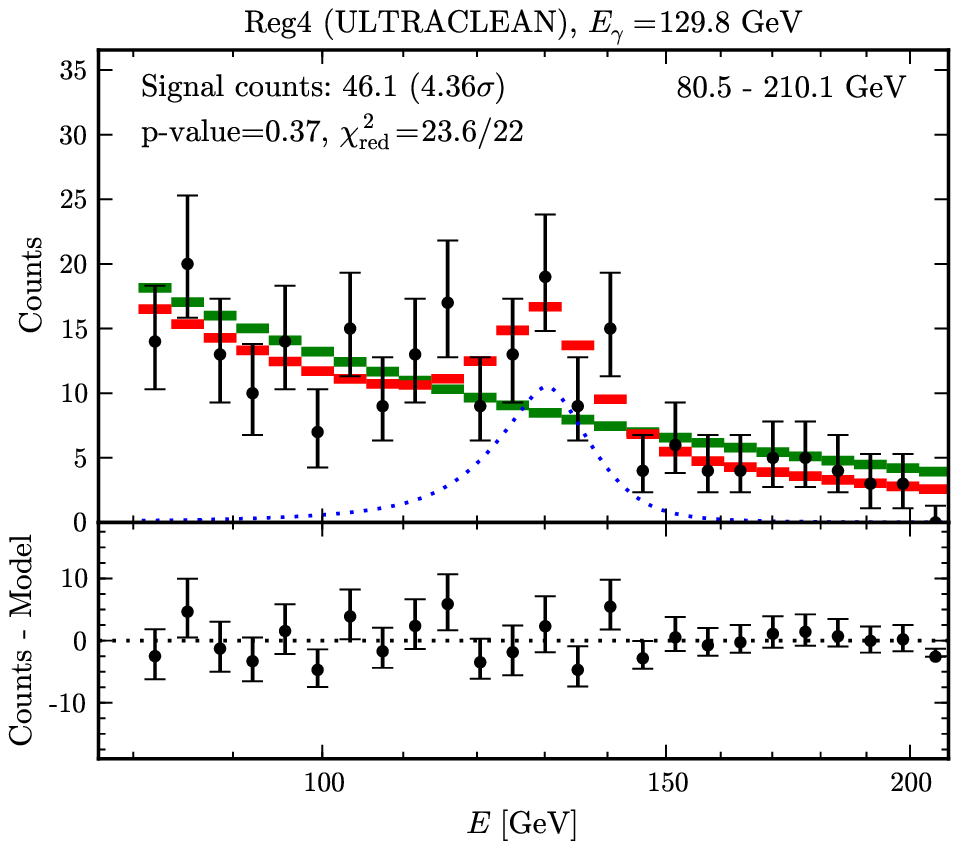}
  \caption{
  \emph{Upper sub-panels:} the measured events with statistical errors are
  plotted in \emph{black}. The \emph{horizontal bars} show the best-fit models
  with (\emph{red}) and without DM (\emph{green}), the \emph{blue dotted
  line} indicates the corresponding line flux component alone. In the \emph{lower
  sub-panel} we show residuals after subtracting the model with line
  contribution. Note that we rebinned the data to fewer bins after performing
  the fits in order to produce the plots and calculate the $p$-value and the
  reduced $\chi_r^2\equiv\chi^2/\text{dof}$. The counts are listed in
  Tabs.~\ref{tab:tablesReg2}, \ref{tab:tablesReg3} and \ref{tab:tablesReg4}.}
  \label{fig:spectra}
\end{figure}

The fits that yield the highest significance for a line contribution are shown
in Fig.~\ref{fig:spectra} for the regions Reg2, Reg3 and Reg4, and for SOURCE
and ULTRACLEAN events. In the upper sub-panels, we plot the LAT data with
statistical error bars, as well as the total predicted counts from the
best-fit models with (red bars) and without a gamma-ray line contribution
(green bars). The blue dotted line shows the line flux component alone (before
averaging over the energy bins).
Note that, in
order to improve the readability of the plots and to calculate the indicated
p-values and the reduced $\chi^2_\text{red}$, we rebinned the data to five
times fewer bins than actually used in the spectral fits.\footnote{In
Fig.~\ref{fig:spectra}, we omitted incomplete bins at the right end of the
energy window. When calculating $\chi^2_\text{red}$, we use the c-statistic
$\sum_i2(\mu_i-c_i)+2c_i\log(c_i/\mu_i)$.} The lower sub-panel shows the count
residuals after subtracting the model with line.  In most of the regions, the
spectral signature that is responsible for the large TS values can be easily
recognized by eye. The number of signal events ranges between 46 and 88, the
statistical significance between $2.8\sigma$ and $4.6\sigma$; the p-values and
residual plots confirm that the fits to the data are reasonable and do not
exhibit systematic discrepancies at low or high energies.\medskip

If we interpret the observed signature as being due to DM annihilation into a
photon pair via $\chi\chi\to\gamma\gamma$, we can constrain the DM mass
$m_\chi$ (which then just equals the line energy, $m_\chi=E_\gamma$) and the
partial annihilation cross-section $\langle\sigma
v\rangle_{\chi\chi\to\gamma\gamma}$.  The corresponding values for processes
like $\chi\chi\to\gamma Z, \gamma h$ follow from a straightforward
rescaling~\cite{Vertongen:2011mu}. The inset of Fig.~\ref{fig:sig} shows a
zoom into the most interesting region of the TS plot.\footnote{To generate the
inset, we did \emph{not} use sliding energy windows but kept the position of
the energy window fixed at the position that corresponds to the $E_\gamma$
with the largest TS.} From there, one can read off the DM mass that best fits
the data together with its error bars.  From the region with the largest TS
value, Reg4 SOURCE class, we obtain $m_\chi=129.8\pm 2.4^{+7}_{-13}\GeV$. The
indicated errors are respectively statistical and systematical, the latter
being due to uncertainties in the overall energy calibration of the LAT,
$\Delta E/E= ^{+5\%}_{-10\%}$~\cite{Fermi:caveatsPASS7}. 

\begin{figure}
  \centering
  \includegraphics[width=.49\linewidth]{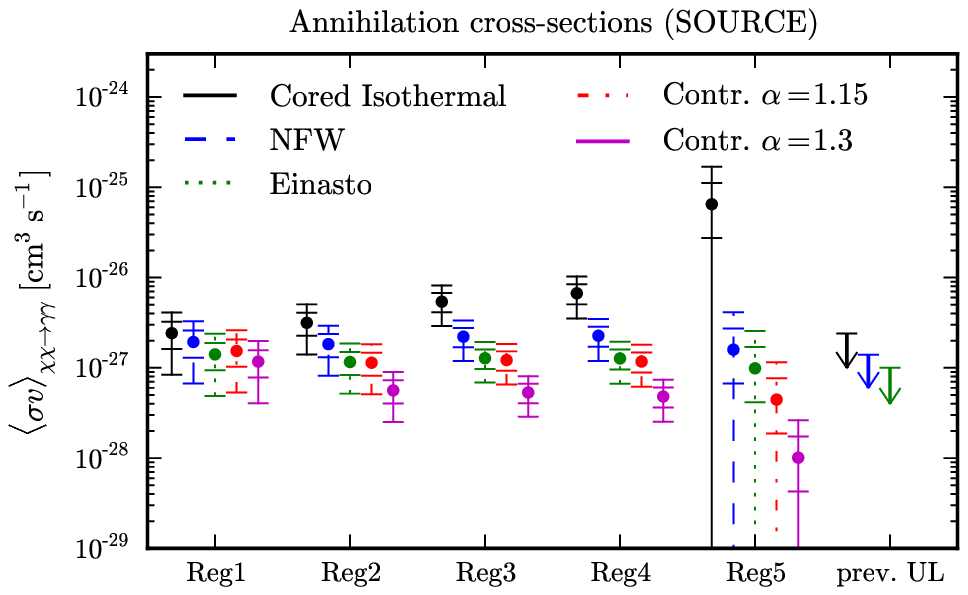}
  \includegraphics[width=.49\linewidth]{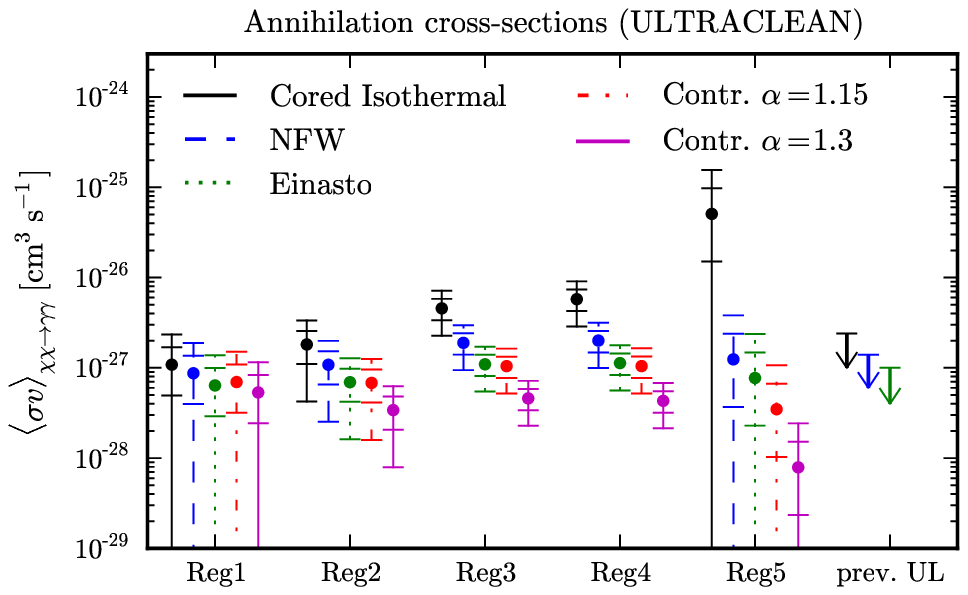}
  \caption{Best-fit values for the annihilation cross-section into a photon
  pair, as obtained for different DM halo profiles in the different target
  regions of Fig.~\ref{fig:reg}, together with their 68.2\%~CL and
  95.5\%~CL errors. Previous upper 95\%~CL limits are shown for
  comparison~\cite{Fermi:Symp2011Talk, Edmonds:2011PhD}. All values are
  derived assuming that $m_\chi=129.0\GeV$.}
  \label{fig:cs}
\end{figure}

In Fig.~\ref{fig:cs} we show central values and $68.2\%$~CL and $95.5\%$~CL
errors for the annihilation cross-section $\langle\sigma
v\rangle_{\chi\chi\to\gamma\gamma}$ as derived from Reg1 to Reg5, assuming
different DM profiles and $m_\chi=129.0\GeV$. We stress that the DM profiles
only affect how signal normalizations from the spectral fits translate into
annihilation cross-sections, but \emph{not} the actual spectral analysis
itself.  In most of the regions, non-zero values of the annihilation
cross-section are preferred at more than $95.5\%$~CL, in agreement with
Fig.~\ref{fig:sig}.  In case of the cored isothermal or the contracted DM
profiles, we find some tension between the annihilation cross-sections
obtained from different target regions.  In case of the Einasto and NFW
profiles, however, the values are mutually consistent. Using SOURCE class
events and the spectrum from Reg4, we find best-fit annihilation
cross-sections
of $\langle\sigma v\rangle_{\chi\chi\to\gamma\gamma}
=(1.27\pm0.32^{+0.18}_{-0.28})\times 10^{-27}\cm^3\s^{-1}$ in case of the
Einasto profile, and of $\langle\sigma v\rangle_{\chi\chi\to\gamma\gamma}
=(2.27\pm0.57^{+0.32}_{-0.51})\times 10^{-27}\cm^3\s^{-1}$ in case of the NFW
profile. The systematic uncertainties are here derived from the effective area
(about $10\%$~\cite{Fermi:caveatsPASS7}) and from the energy calibration.
\medskip

\begin{figure}
  \centering
  \includegraphics[width=.49\linewidth]{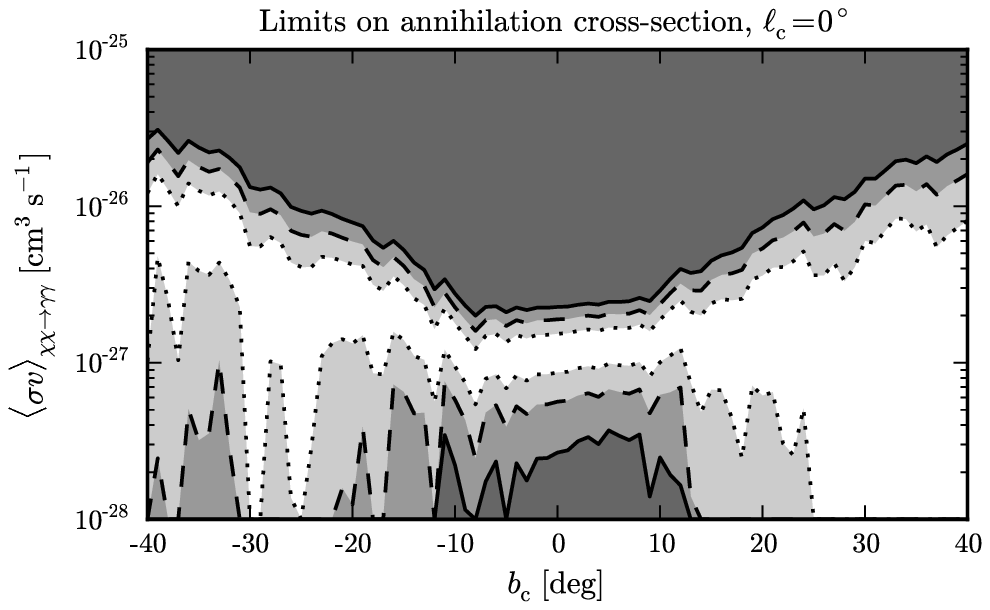}
  \includegraphics[width=.49\linewidth]{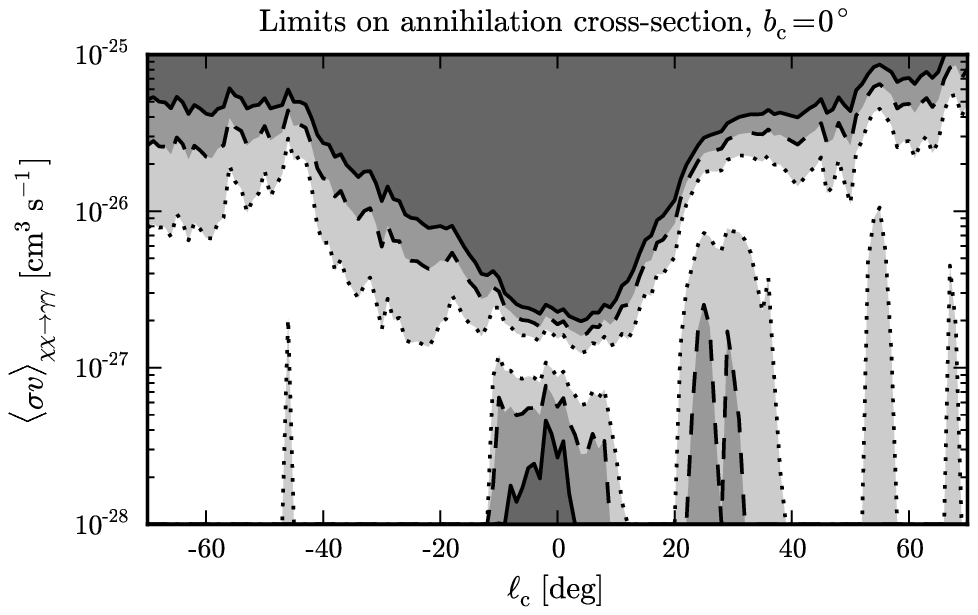}
  \caption{$68.2\%$ (\emph{dotted}), $95.5\%$ (\emph{dashed}) and $99.7\%$~CL
  (\emph{solid}) band of the annihilation cross-section $\langle \sigma
  v\rangle_{\chi\chi\to\gamma\gamma}$ obtained when using a circular target
  region of radius $10^\circ$ centered at the indicated values of
  $\ell_\text{c}$ and $b_\text{c}$. The \emph{left} (\emph{right}) panel shows
  a latitudinal (longitudinal) scan. We assumed $m_\chi=129.0\GeV$ during the
  fits. Note that the abscissa scales are different in both plots.}
  \label{fig:scan}
\end{figure}

In order to test the locality of the observed signature, we extract the
gamma-ray energy spectra from a large number of circular target regions with a
radius of $10^\circ$. These regions are either centered along the Galactic
disk with $b_c=0^\circ$, or they are centered at $\ell_c=0^\circ$ from the
Galactic north to south pole; $\ell_c$ and $b_c$ denote the central
coordinates of the target regions.\footnote{Note that for $b_c\neq0^\circ$
these regions are circular with respect to the projection used in
Fig.~\ref{fig:reg}.} From each of these target regions, we derived the
$68.2\%$, $95.5\%$ and $99.7\%$~CL upper- and lower-limits on the annihilation
cross-section $\langle\sigma v\rangle_{\chi\chi\to\gamma\gamma}$, assuming
$E_\gamma=129.0\GeV$ and using the SOURCE event class. As shown in
Fig.~\ref{fig:scan}, we find that at $99.7\%$~CL non-zero values of the
annihilation cross-section are only preferred close to the GC; the observed
signature disappears when moving to larger values of $|\ell_c|$ or $|b_c|$. 
\medskip

\begin{figure}
  \centering
  \includegraphics[width=.49\linewidth]{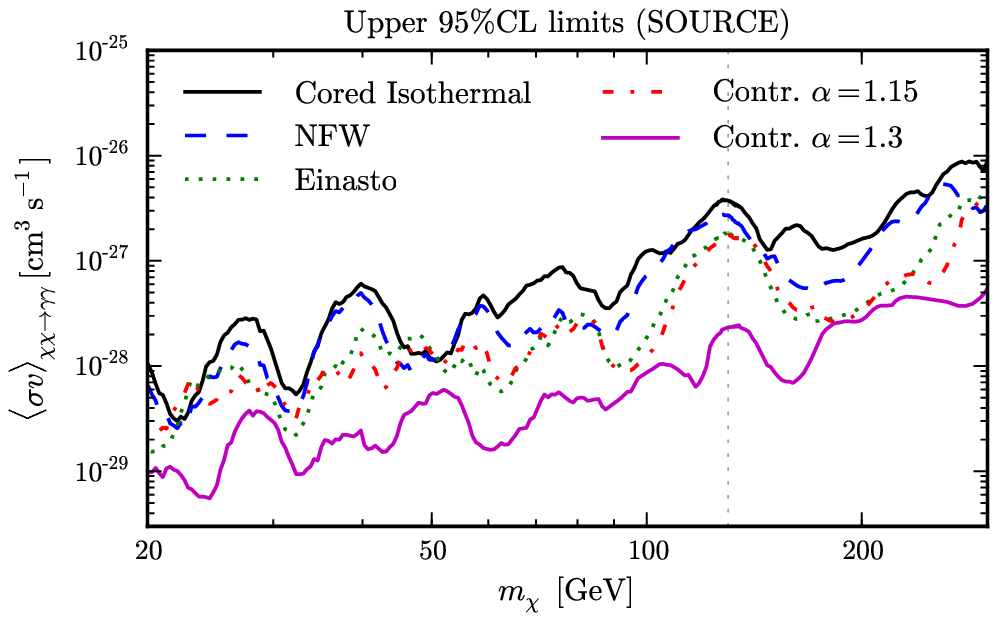}
  \includegraphics[width=.49\linewidth]{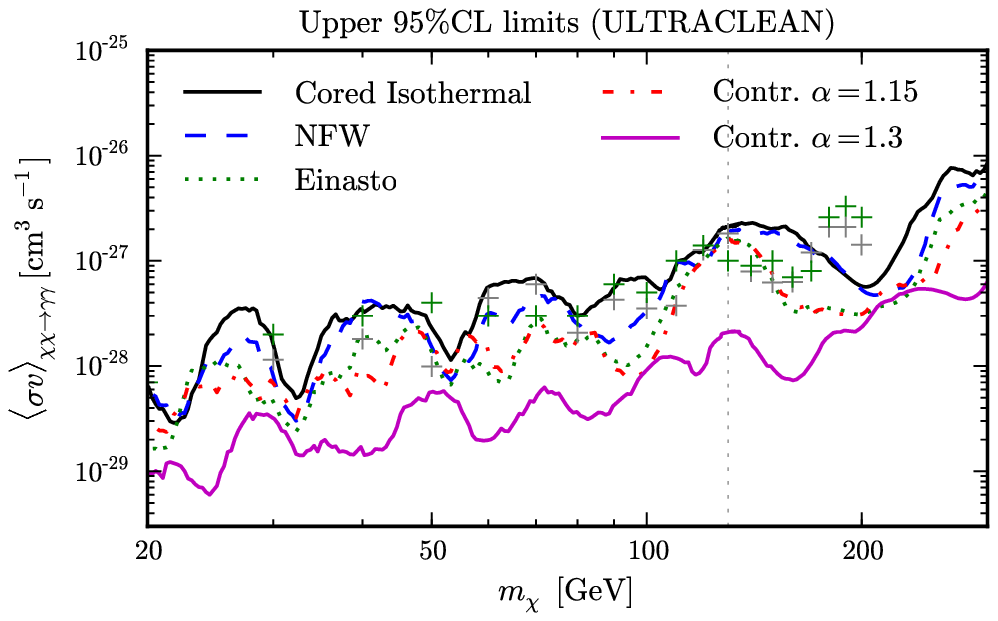}
  \caption{Upper 95\%~CL limits on the annihilation cross-section for
  $\chi\chi\to\gamma\gamma$, for different DM halo profiles, as obtained for
  the SOURCE event class. We used the correspondingly optimized target regions
  from Fig.~\ref{fig:reg}. The green crosses show previous limits from
  Refs.~\cite{Fermi:Symp2011Talk, Edmonds:2011PhD}, assuming an Einasto
  profile; for comparison, the gray crosses show the limits that we obtain
  when using the energy windows and target region from
  Refs.~\cite{Fermi:Symp2011Talk, Edmonds:2011PhD} with 43 months of data. The
  gray dotted line indicates $m_\chi=130\GeV$.}
  \label{fig:lim}
\end{figure}

We show $95\%$~CL upper limits on the annihilation cross-section
$\langle\sigma v\rangle_{\chi\chi\to\gamma\gamma}$ in Fig.~\ref{fig:lim}. For
each DM profile, the limits are derived from the correspondingly optimized
target region. For all considered DM masses, the presented limits are well
below the `thermal cross-section' of $3\times10^{-26}\cm^3\s^{-1}$. As
expected, the limits are weakest in case of the cored isothermal profile, and
strongest for the contracted profile with $\alpha=1.3$. At a mass around
$130\GeV$, they show a pronounced bump that corresponds to the large TS values
in Fig.~\ref{fig:sig}. For the Einasto profile, we plot previous results from
Ref.~\cite{Fermi:Symp2011Talk, Edmonds:2011PhD}---based on two years of data
without a dedicated target region optimization---as green crosses; except for
values around $130\GeV$, we improve these limits by factors of two and more.
\medskip

The signal evidence is based on $46$--$88$ photons, \cf
Fig.~\ref{fig:spectra}. Drawing strong conclusions---in particular on the
spatial distribution or the exact spectral shape of the signature---is thus
hindered by the low number of events collected so far. Furthermore, our
findings could be affected by instrumental systematics, which are difficult to
evaluate with public data and information only. However, we checked that
obvious systematic uncertainties do not significantly affect our findings;
this will be discussed in the next section.

\section{Discussion}
\label{sec:caveats}
\subsection{Instrumental systematics}
\emph{Energy reconstruction and line shape.} 
The present analysis is based on the P7SOURCE\_V6 and P7ULTRACLEAN\_V6 event
classes.  The corresponding publicly available energy information
(CTBBestEnergy) derives from a classification tree analysis of the parametric
correction (PC) and shower profile (SP) reconstruction algorithms.  With
respect to the older event class P6DATACLEAN\_V3, PC and SP were debiased; the
LK (likelihood) reconstruction algorithm, which was found to introduce jagged
peaks that could be mistaken for gamma-ray lines, was removed (see \fex
Ref.~\cite{Edmonds:2011PhD}). The data set underlying the present analysis is
hence expected to be much better suited for line searches than previously
available data.

In previous line searches conducted by the Fermi collaboration (based on the
SP energy alone), the spectral shape of reconstructed gamma-ray lines was
carefully studied by means of Monte Carlo (MC) simulations that were
calibrated with high-energy electron beam tests~\cite{Edmonds:2011PhD}; good
fits to the MC results were obtained by a superposition of three Gaussians,
leading to a peaked line with extended tails. In our analysis, the line shape
is derived from the IRF that ships with the ScienceTools; it has a different
analytical form, is somewhat broader, and related systematic uncertainties are
for us difficult to control. In order to estimate to what extent different
spectral line shapes may affect our results, we took the line shape from
Fig.~5.3 of Ref.~\cite{Edmonds:2011PhD}, as well as a simple single Gaussian
with a FWHM identical to the one derived from the public IRF ($13.6\%$), and
used them in the fits. We find that the TS value corresponding to the observed
signatures in Reg4 (SOURCE) decreases by 0.7 and 3.0, respectively. The
respective best-fit cross-sections are lowered by $8\%$ and $30\%$. A more
careful modeling of the line shape than what we could do in the present work
is hence indeed relevant as it obviously affects the results, but it is
extremely unlikely that it would invalidate our findings.

\emph{Effective area and residual cosmic rays.} Uncertainties in the effective
area of the LAT have a large point-to-point correlation. On scales
corresponding to the width of a gamma-ray line, they are about $2\%$ (and grow
to $15\%$ at larger scales)~\cite{Fermi:caveatsPASS7}. This is much smaller
than the fractional contribution of the observed feature to the overall
flux, and can be hence neglected to good approximation. In any case, a sharp
line-like feature in the effective area (or residual contamination from the
nearly isotropic CRs) would also affect regions of the sky away from the GC.
In light of Fig.~\ref{fig:scan} and the below subsampling analysis, this option
appears very unlikely.

\emph{Event selection.} We checked that the signature appears in both, front-
and back-converted ULTRACLEAN events separately, with a higher significance in
back-converted events. The signature grew over time, with $TS=2.4$ (8.8, 16.9)
when taking only into account only the first 53 (107, 134) weeks of data from
Reg3 SOURCE class. Furthermore, we checked that our results remain
practically unchanged when using the filter cut DATA\_QUAL==1 \&\&
IN\_SAA!=T \&\& LAT\_CONFIG==1 \&\& ABS(ROCK\_ANGLE)$<$52 instead, which
reduces the number of events by $2\%$ with respect to the adopted
DATA\_QUAL==1 cut.
 
\begin{figure}
  \begin{center}
    \includegraphics[width=0.7\linewidth]{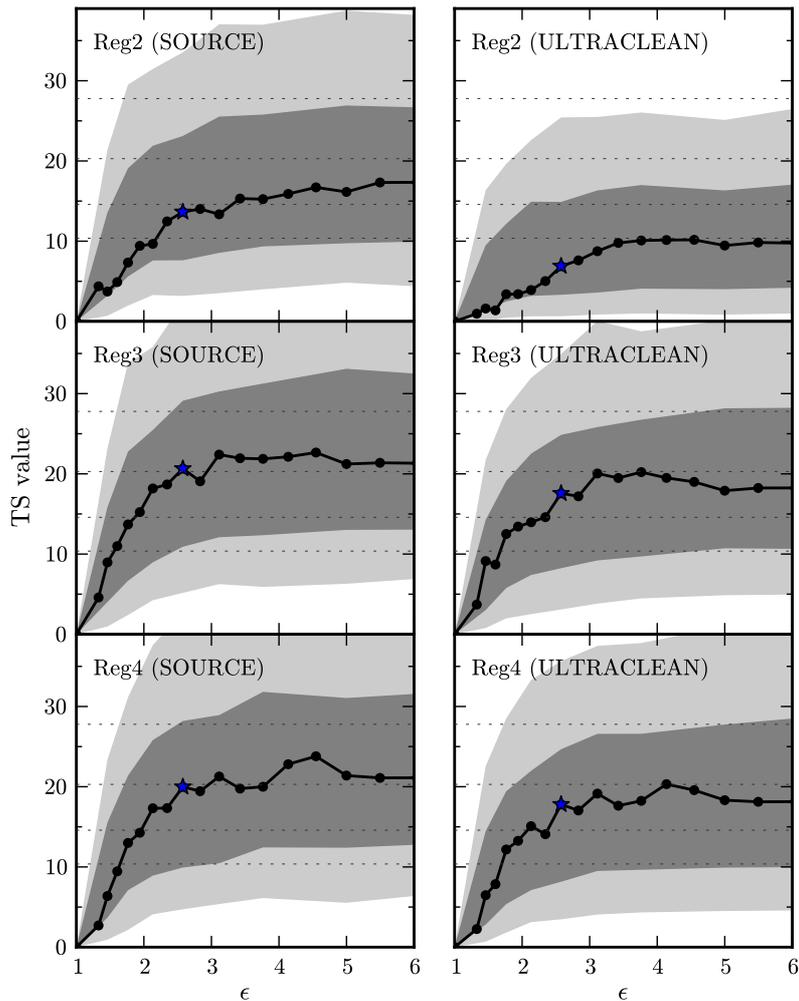}
  \end{center}
  \caption{The \emph{black} line shows how the TS value changes as function of
  the adopted energy window size, $\epsilon$. The \emph{gray shaded} areas are
  the $68.2\%$ and $95.5\%$~CL bands for the TS values obtained from a MC
  simulation. The energy windows borders are calculated according to
  Eq.~\eqref{eqn:ewin}. We assumed $E_\gamma=129.0\GeV$ when generating the
  plot.}
  \label{fig:ewin}
\end{figure}

\subsection{Energy window size} 
As already discussed above, one critical component in the analysis is the
choice of the energy window size.  In Fig.~\ref{fig:ewin}, the black lines
show the TS values obtained when adopting different window sizes $\epsilon$,
assuming a gamma-ray line energy of $E_\gamma=129.0\GeV$. The window borders
are given by Eq.~\eqref{eqn:ewin} above, and we show results for Reg2 to Reg4
and SOURCE and ULTRACLEAN events.  The window size that we actually used
during the main analysis is $\epsilon \simeq 2.58$ and indicated by the blue
star. As one can see from Fig.~\ref{fig:ewin}, an increase of the window size
in reasonable ranges leaves the TS values approximately unchanged.  On the
other hand, a further decrease of the window size would also decreases the TS
value.  This is expected, because the energy range over which the background
normalization is fixed by the data becomes smaller and less constraining in
that case.

To study in more detail how well the behaviour of the TS values follows the
statistical expectations, we perform a simple MC analysis: For each of the six
cases in Fig.~\ref{fig:ewin}, we derive the best-fit model with line from an
energy window with size $\epsilon=6.64$. From this model, we generate 400
mock data sets and refit them using smaller windows. The $68.2\%$~CL and
$95.5\%$~CL bands of the TS values obtained in this way are shown in
Fig.~\ref{fig:ewin} by the dark and light shaded gray areas. We find that the
actual behaviour of the observed TS value as function of $\epsilon$ follows
very well the MC results. This supports our choice of a window size of
$\epsilon\simeq2.58$ at $E_\gamma=129.0\GeV$. Smaller windows would
significantly decrease the statistical power of the analysis.

Lastly, we checked that fixing the slope of the background power-law to
$\gamma=2.6$ (2.5, 2.7) results in TS values of 20.4 (19.4, 21.5) in case of
Reg4 (SOURCE); this again has little impact on our findings. A value of
$\simeq2.6$ is physically expected since the gamma-ray flux in Reg4 is
dominated by diffuse photons from cosmic-ray proton collisions with the
interstellar medium.

\begin{figure}
  \centering
  \includegraphics[width=.65\linewidth]{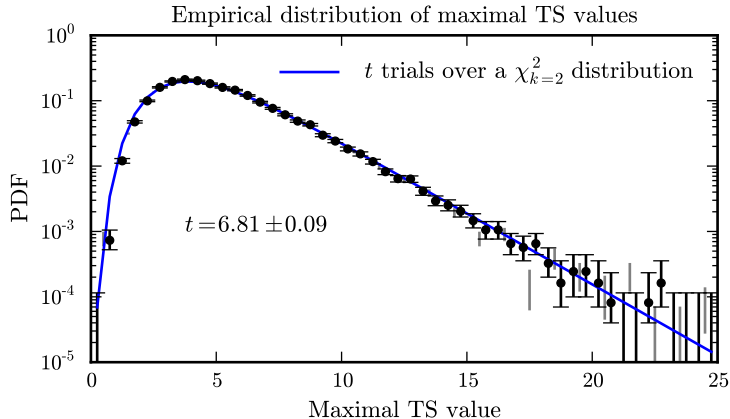}
  \caption{Black: Probability distribution function of maximal TS values that
  we found when performing line searches in 25000 Monte Carlo samples without
  line signal as described in the text. Gray: results obtained from 21000
  subsamples of the hemisphere away from the GC.  The blue line shows the
  fitting function $\text{PDF}_\text{max}(t; x)$; see text for details.}
  \label{fig:hist}
\end{figure}

\subsection{Monte Carlo and subsampling analysis} 
\label{sec:MC}
In order to estimate the statistical significance of the observed excess, we
perform a simple Monte Carlo analysis of our line search. We generate 25000
mock spectra that follow a power-law with slope $-2.3$ (like in Reg1 SOURCE
class), such that each mock data set has the same number of counts as
observed in Reg3 SOURCE. In each of these spectra, we perform the above spectral
analysis, and record the corresponding largest TS value.  The distribution of
these maximal TS values is shown by the black dots in Fig.~\ref{fig:hist} with
statistical error bars. We fit this distribution with the probability
distribution function (PDF) of the maximum from $t$ trials over a
$\chi^2_{k=2}$ distribution, which reads
\begin{align}
  \text{PDF}_\text{max}(t; x) \equiv \frac{d}{dx}\text{CDF}(\chi^2_{k=2};
  x)^t\;,
  \label{eqn:CDF}
\end{align}
leaving $t$ as free parameters in the fit. Note that our $t$ corresponds to
the 'effective number of search regions' discussed in
Ref.~\cite{Gross:2010qma} (related to the number of upcrossings), and $k=2$ is
in our case the theoretical expectation for the tail of the distribution.
Since we are mainly interested in this distribution tail, we only take TS
values above 5 into account in the fit. However, as shown in
Fig.~\ref{fig:hist} by the blue line, we find that $t=6.8$ trials well
reproduce the empirical result over the whole range. We checked that
increasing/decreasing the energy window or sample size within reasonable
values does not change these results.

In order to confirm the above Monte Carlo results with real data, we perform a
subsampling analysis of the LAT data in regions where no signal is expected.
To this end, we exclude all data from the hemisphere towards the GC. The
events from the region away from the GC, $|b|<84^\circ$ and $|\ell|>90^\circ$,
are distributed in energy bins $i$, the number of events in each energy bin is
called $s_i$. From these we generate 21000 test spectra; the expectation value
for each energy bin is given by $\mu_i=0.01s_i$. The small prefactor ensures
that statistical fluctuations within the original data do not destroy the tail
distribution of the subsampling analysis. Like above, we record for each of
the test spectra the largest TS value found. The results are shown in
Fig.~\ref{fig:hist} by the gray bands (which indicate the statistical error
bars). They are in excellent agreement with the Monte Carlo results. We
checked that this agreement does not strongly depend on the energy window size
or on the above prefactor; it holds in particular also in the case where the
number of events per test spectra equals the number of events in Reg3.  

We emphasize that line-like artefacts in the effective area of the LAT or
contamination with isotropic cosmic rays would most likely have shown up as
disagreement between our Monte Carlo and subsampling results, since they
should also affect the fluxes observed towards the anti-galactic center.

\subsection{Comparison with previous results} 
The strongest previously presented upper limits on gamma-ray lines at an
energy of $E_\gamma=130\GeV$ read $\langle\sigma
v\rangle_{\chi\chi\to\gamma\gamma}<1.4\ (1.0,\ 2.4)\times
10^{-27}\cm^3\s^{-1}$ for an NFW (Einasto, cored isothermal) DM profile (at
$95\%$~CL)~\cite{Fermi:Symp2011Talk, Edmonds:2011PhD}, and are based on two
years of data and an analysis of a large sky region. In case of the Einasto
profile, these limits are marginally consistent with the different best-fit
cross-section that we show in Fig.~\ref{fig:cs}. In case of the NFW (cored
isothermal) profile, however, there is a two sigma tension between previous
limits and the best-fit value from Reg4. This could be read as a weak
preference for the Einasto (or for a compressed) DM profile. However, it could
also indicate systematics related \fex to the shape of the gamma-ray line that
biases the obtained cross-sections to high, or it could be due to statistical
fluctuations in the different underlying data sets. In any case, these limits
would set interesting and complementary constraints on the allowed DM profiles
if the 130 GeV signature is due to DM annihilation.

In Refs.~\cite{Fermi:Symp2011Talk, Edmonds:2011PhD}, which is based on the SP
energy only, the significance for a signal at $E_\gamma=130\GeV$ was found to
be $\text{TS}<1$. As a test whether this conflicts with our results, we
implemented the target region and energy windows from
Refs.~\cite{Fermi:Symp2011Talk, Edmonds:2011PhD} (without doing a point-source
subtraction), and repeated their scan over the discrete line energies
$E_\gamma=30, 40, \dots 200\GeV$, using 43 months of ULTRACLEAN data. In this
way, we find the TS values (upper limits) that are shown by the gray crosses
in the right panel of Fig.~\ref{fig:sig} (Fig.~\ref{fig:lim}). Indeed, no
strong indication for a signature at 130 GeV appears, whereas the resulting
limits are in good agreement with---and slightly stronger than---the two-years
results from~\cite{Fermi:Symp2011Talk, Edmonds:2011PhD}. This illustrates that
a continuous scan over of gamma-ray line energies, a careful selection of
energy windows and an improved selection of target regions is extremely
important when searching for very faint signatures.

\section{Conclusions}
\label{sec:conclusions}

We presented a refined search for gamma-ray lines with the Fermi Large Area
Telescope. The main improvement with respect to previous analyses is the use
of a new optimization algorithm that automatically selects target regions with
the largest signal-to-noise ratio, depending on the adopted Galactic dark
matter profile and the LAT data. Besides, we updated previous results to 43
months of data and used the most recent public event selection. 

We found a $4.6\sigma$ indication for the presence of a gamma-ray line at
$E_\gamma\approx130\GeV$.\footnote{We reported this signature before in
Ref.~\cite{Bringmann:2012vr} in context of internal Bremsstrahlung signals
from DM.} Taking into account the look-elsewhere effect, the significance is
$3.2\sigma$. By scanning through different regions of the sky, we find that
the signature appears close to the Galactic center only. If interpreted in
terms of dark matter annihilation into photon pairs,
$\chi\chi\to\gamma\gamma$, the observations constrain the dark matter mass to
$m_\chi=129.8\pm 2.4^{+7}_{-13}\GeV$. The annihilation cross-sections derived
from different target regions are consistent with each other in the case of
the Einasto or NFW dark matter profiles. For these, we obtain best-fit
annihilation cross-sections of $\langle\sigma
v\rangle_{\chi\chi\to\gamma\gamma}=
(1.27\pm0.32^{+0.18}_{-0.28})\times10^{-27}\cm^3\s^{-1}$ and $\langle\sigma
v\rangle_{\chi\chi\to\gamma\gamma}=
(2.27\pm0.57^{+0.32}_{-0.51})\times10^{-27}\cm^3\s^{-1}$, respectively.
Assuming a thermal relic, this corresponds to a branching ratio into
$\gamma\gamma$ final states of $4$--$8\%$, which is much larger than what the
generic one-loop suppression suggests. We examined our statistical method with
Monte Carlo methods and with a subsampling analysis of the LAT data in the
$|\ell|>90^\circ$ hemisphere, and
find good agreement with theoretical expectations. We performed several
checks to exclude that obvious systematics of the LAT could invalidate our
findings. In particular the fact that the signal peaks around the Galactic
center makes a purely instrumental cause very unlikely. 

The observation of a gamma-ray line from dark matter annihilation in the data
of the Fermi LAT is a fascinating possibility. However, a few caveats are in
order. First, the presented analysis is based on publicly available data and
information only, hence we cannot take into account all possible instrumental
effects.  Second, the signal evidence is right now based on around $50$
photons; it will require a few years of more data to settle its existence on
statistical grounds. Third, due to the low number of events it is difficult to
study in detail the spatial and spectral characteristics of the signature,
which might still leave room for some exotic astrophysical explanation.

In any case, we showed that a careful selection of target regions is extremely
important when searching for gamma-ray lines. We hope that our work motivates a
refined search for such features in the Fermi LAT data, including all
instrumental effects, an improved event selection \emph{and} a target
region optimization as proposed in this paper.

\paragraph*{Acknowledgments.}
I am very grateful to Torsten Bringmann, Ilias Cholis, Michael Gustafsson,
Dieter Horns, Luca Maccione, David Paneque, Maksim Pshirkov, Georg Raffelt,
Javier Redondo and Stefan Vogl for useful comments on the manuscript and
valuable discussion about the Fermi LAT and statistical methods. I
acknowledges partial support from the European 1231 Union FP7 ITN INVISIBLES
(Marie Curie Actions, PITN-GA-2011-289442). This work makes use of
SciPy~\cite{SciPy},
PyFITS\footnote{\url{http://www.stsci.edu/resources/software_hardware/pyfits}},
PyMinuit\footnote{\url{http://code.google.com/p/pyminuit}} and
IPython~\cite{IPython}.

\paragraph*{Note added.} After submission of this manuscript, the two-years
line search of the LAT collaboration was released~\cite{Ackermann:2012qk}. The
results were already presented on different conferences during last
year~\cite{Fermi:Symp2011Talk}, details can be found in
Ref.~\cite{Edmonds:2011PhD}. We comment on them where appropriate.

\bibliography{}
\bibliographystyle{JHEP}

\appendix
\section{Event Tables}
\label{apx:tables}

\begin{table}[h]
  \scriptsize
  \centering
  \begin{tabular}{cccc}
    \toprule
    $E$ bins & \#Counts & \#Signal & Flux \\
    $\rm [GeV]$ &  &  & $\rm [ph/cm^2/s/sr]$ \\\midrule
    \input{Reg2SOURCE.tab}
    \bottomrule
  \end{tabular}
  \hspace{0.3cm}
  \begin{tabular}{cccc}
    \toprule
    $E$ bins & \#Counts & \#Signal & Flux \\
    $\rm [GeV]$ &  &  & $\rm [ph/cm^2/s/sr]$ \\\midrule
    \input{Reg2ULTRACLEAN.tab}
    \bottomrule
  \end{tabular}
  \caption{From left to right, the tables show the energy bins, the number
  of observed counts, the best-fit signal counts, and the observed fluxes that
  appear in Fig.~\ref{fig:spectra}. The left table corresponds to Reg2 SOURCE
  class, the right table to Reg2 ULTRACLEAN class. Although in our main
  analysis we use five times smaller energy bins, the listed numbers are
  enough to reproduce our results with good accuracy.}
  \label{tab:tablesReg2}
\end{table}

\begin{table}[h]
  \centering
  \scriptsize
  \begin{tabular}{cccc}
    \toprule
    $E$ bins & \#Counts & \#Signal & Flux \\
    $\rm [GeV]$ &  &  & $\rm [ph/cm^2/s/sr]$ \\\midrule
    \input{Reg3SOURCE.tab}
    \bottomrule
  \end{tabular}
  \hspace{0.3cm}
  \begin{tabular}{cccc}
    \toprule
    $E$ bins & \#Counts & \#Signal & Flux \\
    $\rm [GeV]$ &  &  & $\rm [ph/cm^2/s/sr]$ \\\midrule
    \input{Reg3ULTRACLEAN.tab}
    \bottomrule
  \end{tabular}
  \caption{Same as Tab.~\ref{tab:tablesReg2}, but for Reg3.}
  \label{tab:tablesReg3}
\end{table}

\begin{table}[h]
  \centering
  \scriptsize
  \begin{tabular}{cccc}
    \toprule
    $E$ bins & \#Counts & \#Signal & Flux \\
    $\rm [GeV]$ &  &  & $\rm [ph/cm^2/s/sr]$ \\\midrule
    \input{Reg4SOURCE.tab}
    \bottomrule
  \end{tabular}
  \hspace{0.3cm}
  \begin{tabular}{cccc}
    \toprule
    $E$ bins & \#Counts & \#Signal & Flux \\
    $\rm [GeV]$ &  &  & $\rm [ph/cm^2/s/sr]$ \\\midrule
    \input{Reg4ULTRACLEAN.tab}
    \bottomrule
  \end{tabular}
  \caption{Same as Tab.~\ref{tab:tablesReg2}, but for Reg4.}
  \label{tab:tablesReg4}
\end{table}

\end{document}

%% file: Reg2SOURCE.tab
79.3--82.4 & 64 & 0.34 & 1.37e-10 \\
82.4--85.6 & 84 & 0.42 & 1.73e-10 \\
85.6--89.0 & 74 & 0.53 & 1.47e-10 \\
89.0--92.5 & 66 & 0.68 & 1.27e-10 \\
92.5--96.1 & 63 & 0.88 & 1.17e-10 \\
96.1--99.8 & 68 & 1.17 & 1.22e-10 \\
99.8--103.7 & 68 & 1.61 & 1.18e-10 \\
103.7--107.8 & 65 & 2.31 & 1.09e-10 \\
107.8--112.0 & 55 & 3.54 & 8.88e-11 \\
112.0--116.4 & 72 & 5.81 & 1.12e-10 \\
116.4--120.9 & 50 & 9.68 & 7.53e-11 \\
120.9--125.7 & 46 & 15.18 & 6.7e-11 \\
125.7--130.6 & 67 & 18.89 & 9.43e-11 \\
130.6--135.7 & 51 & 13.76 & 6.94e-11 \\
135.7--141.0 & 36 & 7.07 & 4.74e-11 \\
141.0--146.5 & 35 & 3.04 & 4.45e-11 \\
146.5--152.2 & 34 & 1.30 & 4.18e-11 \\
152.2--158.2 & 34 & 0.61 & 4.04e-11 \\
158.2--164.4 & 28 & 0.34 & 3.22e-11 \\
164.4--170.8 & 23 & 0.20 & 2.56e-11 \\
170.8--177.5 & 26 & 0.12 & 2.79e-11 \\
177.5--184.4 & 22 & 0.07 & 2.29e-11 \\
184.4--191.6 & 19 & 0.04 & 1.91e-11 \\
191.6--199.1 & 14 & 0.03 & 1.36e-11 \\

%% file: Reg2ULTRACLEAN.tab
81.1--84.3 & 47 & 0.16 & 9.39e-11 \\
84.3--87.6 & 51 & 0.20 & 9.83e-11 \\
87.6--91.0 & 48 & 0.25 & 8.94e-11 \\
91.0--94.6 & 44 & 0.33 & 7.91e-11 \\
94.6--98.3 & 41 & 0.43 & 7.12e-11 \\
98.3--102.1 & 39 & 0.58 & 6.55e-11 \\
102.1--106.1 & 42 & 0.81 & 6.8e-11 \\
106.1--110.3 & 31 & 1.16 & 4.84e-11 \\
110.3--114.6 & 49 & 1.78 & 7.37e-11 \\
114.6--119.1 & 36 & 2.94 & 5.23e-11 \\
119.1--123.7 & 20 & 4.96 & 2.81e-11 \\
123.7--128.6 & 36 & 7.99 & 4.9e-11 \\
128.6--133.6 & 41 & 10.57 & 5.39e-11 \\
133.6--138.8 & 25 & 8.16 & 3.18e-11 \\
138.8--144.3 & 24 & 4.25 & 2.95e-11 \\
144.3--149.9 & 22 & 1.82 & 2.62e-11 \\
149.9--155.8 & 21 & 0.76 & 2.41e-11 \\
155.8--161.9 & 25 & 0.35 & 2.77e-11 \\
161.9--168.2 & 13 & 0.18 & 1.39e-11 \\
168.2--174.8 & 15 & 0.10 & 1.55e-11 \\
174.8--181.6 & 16 & 0.06 & 1.59e-11 \\
181.6--188.7 & 14 & 0.04 & 1.35e-11 \\
188.7--196.1 & 11 & 0.02 & 1.03e-11 \\
196.1--203.8 & 10 & 0.01 & 9.02e-12 \\
203.8--211.7 & 6 & 0.01 & 5.24e-12 \\

%% file: Reg3SOURCE.tab
80.5--83.7 & 31 & 0.26 & 2.78e-10 \\
83.7--86.9 & 33 & 0.33 & 2.86e-10 \\
86.9--90.3 & 25 & 0.41 & 2.09e-10 \\
90.3--93.9 & 21 & 0.53 & 1.69e-10 \\
93.9--97.6 & 32 & 0.69 & 2.49e-10 \\
97.6--101.4 & 24 & 0.91 & 1.8e-10 \\
101.4--105.3 & 31 & 1.26 & 2.25e-10 \\
105.3--109.5 & 17 & 1.80 & 1.19e-10 \\
109.5--113.7 & 26 & 2.76 & 1.76e-10 \\
113.7--118.2 & 32 & 4.53 & 2.09e-10 \\
118.2--122.8 & 21 & 7.56 & 1.33e-10 \\
122.8--127.6 & 23 & 11.87 & 1.41e-10 \\
127.6--132.6 & 32 & 14.82 & 1.89e-10 \\
132.6--137.8 & 21 & 10.83 & 1.2e-10 \\
137.8--143.2 & 22 & 5.58 & 1.21e-10 \\
143.2--148.8 & 7 & 2.40 & 3.74e-11 \\
148.8--154.6 & 12 & 1.03 & 6.19e-11 \\
154.6--160.6 & 9 & 0.48 & 4.49e-11 \\
160.6--166.9 & 9 & 0.27 & 4.34e-11 \\
166.9--173.4 & 10 & 0.16 & 4.66e-11 \\
173.4--180.2 & 10 & 0.09 & 4.51e-11 \\
180.2--187.3 & 8 & 0.06 & 3.48e-11 \\
187.3--194.6 & 8 & 0.03 & 3.37e-11 \\
194.6--202.2 & 6 & 0.02 & 2.44e-11 \\

%% file: Reg3ULTRACLEAN.tab
80.5--83.7 & 21 & 0.18 & 2.03e-10 \\
83.7--86.9 & 23 & 0.23 & 2.14e-10 \\
86.9--90.3 & 21 & 0.29 & 1.89e-10 \\
90.3--93.9 & 18 & 0.37 & 1.56e-10 \\
93.9--97.6 & 21 & 0.49 & 1.76e-10 \\
97.6--101.4 & 16 & 0.66 & 1.3e-10 \\
101.4--105.3 & 24 & 0.92 & 1.88e-10 \\
105.3--109.5 & 14 & 1.34 & 1.06e-10 \\
109.5--113.7 & 21 & 2.06 & 1.53e-10 \\
113.7--118.2 & 20 & 3.41 & 1.4e-10 \\
118.2--122.8 & 14 & 5.76 & 9.49e-11 \\
122.8--127.6 & 17 & 9.21 & 1.11e-10 \\
127.6--132.6 & 26 & 11.80 & 1.65e-10 \\
132.6--137.8 & 15 & 8.72 & 9.2e-11 \\
137.8--143.2 & 17 & 4.45 & 1.01e-10 \\
143.2--148.8 & 6 & 1.89 & 3.44e-11 \\
148.8--154.6 & 9 & 0.79 & 4.98e-11 \\
154.6--160.6 & 5 & 0.36 & 2.67e-11 \\
160.6--166.9 & 7 & 0.19 & 3.61e-11 \\
166.9--173.4 & 7 & 0.11 & 3.48e-11 \\
173.4--180.2 & 8 & 0.07 & 3.84e-11 \\
180.2--187.3 & 7 & 0.04 & 3.25e-11 \\
187.3--194.6 & 5 & 0.02 & 2.25e-11 \\
194.6--202.2 & 4 & 0.01 & 1.74e-11 \\

%% file: Reg4SOURCE.tab
80.5--83.7 & 19 & 0.22 & 2.99e-10 \\
83.7--86.9 & 26 & 0.27 & 3.95e-10 \\
86.9--90.3 & 16 & 0.34 & 2.35e-10 \\
90.3--93.9 & 13 & 0.43 & 1.84e-10 \\
93.9--97.6 & 21 & 0.56 & 2.87e-10 \\
97.6--101.4 & 13 & 0.74 & 1.72e-10 \\
101.4--105.3 & 18 & 1.01 & 2.29e-10 \\
105.3--109.5 & 11 & 1.45 & 1.35e-10 \\
109.5--113.7 & 17 & 2.21 & 2.02e-10 \\
113.7--118.2 & 23 & 3.60 & 2.64e-10 \\
118.2--122.8 & 13 & 6.02 & 1.44e-10 \\
122.8--127.6 & 15 & 9.52 & 1.61e-10 \\
127.6--132.6 & 24 & 12.29 & 2.49e-10 \\
132.6--137.8 & 13 & 9.37 & 1.3e-10 \\
137.8--143.2 & 19 & 4.93 & 1.84e-10 \\
143.2--148.8 & 5 & 2.14 & 4.68e-11 \\
148.8--154.6 & 7 & 0.91 & 6.34e-11 \\
154.6--160.6 & 7 & 0.43 & 6.12e-11 \\
160.6--166.9 & 5 & 0.23 & 4.23e-11 \\
166.9--173.4 & 6 & 0.13 & 4.91e-11 \\
173.4--180.2 & 7 & 0.08 & 5.53e-11 \\
180.2--187.3 & 5 & 0.05 & 3.82e-11 \\
187.3--194.6 & 3 & 0.03 & 2.22e-11 \\
194.6--202.2 & 4 & 0.02 & 2.86e-11 \\
202.2--210.1 & 2 & 0.01 & 1.38e-11 \\

%% file: Reg4ULTRACLEAN.tab
80.5--83.7 & 14 & 0.15 & 2.31e-10 \\
83.7--86.9 & 20 & 0.19 & 3.18e-10 \\
86.9--90.3 & 13 & 0.25 & 1.99e-10 \\
90.3--93.9 & 10 & 0.32 & 1.48e-10 \\
93.9--97.6 & 14 & 0.42 & 2e-10 \\
97.6--101.4 & 7 & 0.57 & 9.68e-11 \\
101.4--105.3 & 15 & 0.79 & 2e-10 \\
105.3--109.5 & 9 & 1.14 & 1.16e-10 \\
109.5--113.7 & 13 & 1.75 & 1.61e-10 \\
113.7--118.2 & 17 & 2.88 & 2.04e-10 \\
118.2--122.8 & 9 & 4.87 & 1.04e-10 \\
122.8--127.6 & 13 & 7.81 & 1.46e-10 \\
127.6--132.6 & 19 & 10.18 & 2.06e-10 \\
132.6--137.8 & 9 & 7.68 & 9.42e-11 \\
137.8--143.2 & 15 & 3.96 & 1.52e-10 \\
143.2--148.8 & 4 & 1.69 & 3.92e-11 \\
148.8--154.6 & 6 & 0.70 & 5.67e-11 \\
154.6--160.6 & 4 & 0.32 & 3.65e-11 \\
160.6--166.9 & 4 & 0.17 & 3.52e-11 \\
166.9--173.4 & 5 & 0.10 & 4.24e-11 \\
173.4--180.2 & 5 & 0.06 & 4.09e-11 \\
180.2--187.3 & 4 & 0.03 & 3.17e-11 \\
187.3--194.6 & 3 & 0.02 & 2.3e-11 \\
194.6--202.2 & 3 & 0.01 & 2.22e-11 \\
202.2--210.1 & 0 & 0.01 & 0 \\